\documentclass[twocolumn]{article}

\usepackage{geometry}
\usepackage[dvipsnames]{xcolor}
\usepackage{soul}

\usepackage{subcaption}
\usepackage{booktabs}
\usepackage{adjustbox}
\usepackage{multirow}
\usepackage[all]{nowidow}
\usepackage{amsmath}
\usepackage{amssymb}
\usepackage{bm}
\usepackage{nicefrac}
\usepackage{mathtools}

\usepackage[separate-uncertainty=true,parse-numbers=true]{siunitx}

\usepackage{pgfplots}
\usepackage{tikz}
\pgfplotsset{compat = newest}
\usetikzlibrary{quotes,angles,calc,arrows,patterns,quotes,external}
\usepgfplotslibrary{groupplots}

\usepackage[colorlinks=true, citecolor=blue,urlcolor=blue,linkcolor=blue,filecolor=black]{hyperref}
\usepackage[capitalise,nameinlink,noabbrev]{cleveref} 

\usepackage{authblk}

\usepackage{algorithm}
\usepackage{algpseudocode}

\usepackage{multibib}
\newcites{Supplement}{Supplementary Information References}

\usepackage[T1]{fontenc}
\usepackage[utf8]{inputenc}

\usepackage{nowidow}

\title{A Fiber-pigtailed Quantum Dot Device\\ Generating Indistinguishable Photons at GHz Clock-rates}

\author[1,*]{Lucas Rickert}
\author[2]{Kinga Żołnacz}
\author[1]{Daniel A. Vajner}
\author[1]{Martin von Helversen}
\author[1]{Sven Rodt}
\author[1]{Stephan Reitzenstein}
\author[3,4]{Hanqing Liu}
\author[3,4]{Shulun Li}
\author[3,4]{Haiqiao Ni}
\author[5,6]{Paweł Wyborski}
\author[5]{Grzegorz Sęk}
\author[5]{Anna Musiał}
\author[3,4,*]{Zhichuan Niu}
\author[1,*]{Tobias Heindel}
\affil[1]{Institute of Solid State Physics, Technical University Berlin, Hardenbergstraße 36, 10623 Berlin, Germany}
\affil[2]{Department of Optics and Photonics, Wroclaw University of Science and Technology, Wybrzeże Stanisława Wyspiańskiego 27, 50-370 Wroclaw, Poland}
\affil[3]{Key Laboratory of Optoelectronic Materials and Devices, Institute of Semiconductors, Chinese Academy of Sciences, Beijing 100083, China}
\affil[4]{Center of Materials Science and Optoelectronics Engineering, University of Chinese Academy of Sciences, Beijing 100049, China}
\affil[5]{Department of Experimental Physics, Wroclaw University of Science and Technology, Wybrzeże Stanisława Wyspiańskiego 27, 50-370 Wroclaw, Poland}
\affil[6]{Department of Electrical and Photonics Engineering, Technical University of Denmark, 2800, Kgs. Lyngby, Denmark}
\affil[*]{Corresponding author: lucas.rickert@tu-berlin.de, zcniu@semi.ac.cn, tobias.heindel@tu-berlin.de}
\date{}

\begin{document}


\twocolumn[
\begin{@twocolumnfalse}
\maketitle
\begin{abstract}
\noindent
Solid-state quantum light sources based on semiconductor quantum dots (QDs) are increasingly employed in photonic quantum information applications. Especially when moving towards real-world scenarios outside shielded lab environments, the efficient and robust coupling of nanophotonic devices to single-mode optical fibers offers substantial advantage by enabling "plug-and-play" operation. In this work we present a fiber-pigtailed cavity-enhanced source of flying qubits emitting single indistinguishable photons at clock-rates exceeding $1$\,GHz. This is achieved by employing a fully deterministic technique for fiber-pigtailing optimized QD-devices based on hybrid circular Bragg grating (hCBG) micro-cavities. The fabricated fiber-pigtailed hCBGs feature radiative emission lifetimes of $<80$\,ps, corresponding to a Purcell factor of $\sim$9, a suppression of multi-photon emission events with $g^{(2)}(0)<1$\%, a photon-indistinguishability $>80$\% and a measured single-photon coupling efficiency of 53\% in a high numerical aperture single-mode fiber, corresponding to 1.2~Megaclicks per second at the single-photon detectors under 80\,MHz excitation clock-rates. 
Furthermore, we show that high multi-photon suppression and indistinguishability prevail for excitation clock-rates exceeding $1$\,GHz.
Our results show that Purcell-enhanced
fiber-pigtailed quantum light sources based on hCBG cavities are a prime candidate for applications of quantum information science.
\end{abstract}
\hspace{2cm}
\end{@twocolumnfalse}
]

\section{Introduction}
\label{sec:Intro}

Solid-state quantum emitters~\cite{aharonovich_solid-state_2016,couteau_applications_2023} providing coherent, indistinguishable photons enable scalable photonic quantum technologies~\cite{obrien_photonic_2009}, such as boson sampling~\cite{aaronson_computational_2011}, cluster-state generation~\cite{schwartz_deterministic_2016,istrati_sequential_2020,cogan_deterministic_2023} and device-independent quantum key distribution (QKD) protocols~\cite{acin_device-independent_2007,zhang_device-independent_2022}. Particularly semiconductor quantum dots (QDs) have raised considerable research interests and are currently out-performing other solid-state systems in terms of brightness, multiphoton-suppression and generation of indistinguishable photons~\cite{ding_high-efficiency_2023,tomm_bright_2021,somaschi_near-optimal_2016,schweickert_-demand_2018}.

Two main technological challenges have to be overcome to make QD-based quantum light sources practical for applications outside shielded laboratory environments. Firstly, the integration with compact cryocoolers, required for harnessing their state-of-the art quantum-optical properties and, secondly, the implementation of a robust and efficient fiber-optical interface, enabling the alignment-free harnessing of flying qubits directly in-fiber without any bulk optical components. Conveniently, such a "plug-and-play" QD-quantum light source permanently attached to a single mode fiber (SMF) \cite{xu_plug_2007} is robust enough in presence of vibrations to allow for single-photon emission with the use of compact mechanical cryocoolers~\cite{schlehahn_stand-alone_2018, musial_plugplay_2020}, showing their practicality in first quantum communication experiments~\cite{gao_quantum_2022}.

Fiber-pigtailed structures have been reported based on QDs embedded in photonic structures enhancing the in-fiber photon collection based on geometric effects, such as micro-lenses~\cite{schlehahn_stand-alone_2018} -mesas~\cite{musial_plugplay_2020}, and nano-wires~\cite{cadeddu_fiber-coupled_2016,northeast_optical_2021}. Furthermore, on-chip QD-fiber interfaces have been investigated, which allow transfer of photons emitted by the QD via an evanescent coupling to a fiber in close proximitty~\cite{lee_efficient_2015,daveau_efficient_2017,lee_fiber-integrated_2019}, even with permantent adhesion of fiber and chip~\cite{zeng_cryogenic_2023}.

Plug-and-play sources based on photonic micro-cavities operating deep in the Purcell-enhanced~\cite{purcell_e_m_b10_1946} regime, are of particular interest for advanced quantum optical performance. Here, the reduced radiative lifetime $T_1$ of the embedded emitter allows for higher degrees of coherence, even in the presence of inhomogeneous broadening caused by fluctuating charge environments due to etched surfaces in the QD's vicinity~\cite{liu_high_2018} or coupling to phonons. The resilience against phonon interactions additionally enhances significantly the photon-indistinguishability at elevated temperatures~\cite{grange_reducing_2017,brash_nanocavity_2023}. Cavity-enhanced fiber-pigtailed QD devices thus promise superior performance for advanced applications in quantum information science.

Only few works on cavity-enhanced fiber-pigtailed QD sources are reported in literature so far, typically employing micropillar cavities~\cite{snijders_fiber-coupled_2018,chen_fiber_2021}. The Purcell enhancement achieved in combination with coherent excitation methods in Ref.~\cite{snijders_fiber-coupled_2018} enabled highly indistinguishable photons directly in fiber.     

Another cavity-type that gained research-interest recently is the hybrid circular Bragg grating (hCBG) cavity~\cite{yao_design_2018}, promising considerable Purcell-enhancement and more broadband photon collection efficiencies compared to e.g. micropillar cavities. High potential for fiber-pigtailing these hCBG cavities has been reported based on simulations~\cite{rickert_optimized_2019,barbiero_design_2022,bremer_numerical_2022}, but so far experimental realizations of such a fiber-pigtailed QD based CBG cavity~\cite{jeon_plugandplay_2022} were limited in brightness and Purcell enhancement, and no quantum optical performance beyond the single-photon emission properties was investigated.
 
In this work, we report a fully deterministic fiber-pigtailed quantum light source based on QD-hCBG cavities exhibiting $T_1$-times as low as 76\,ps, corresponding to a Purcell factor close to 9, which allows for operation at GHz clock-rates. We observe a strong suppression of multi-photon emission events associated with $g^{(2)}(0)=0.007(2)$ at 80\,MHz excitation rate and pulsed two-photon indistinguishabilities of up to 82(4)\% (79(4)\%) for 2\,ns (12.5\,ns) temporal separation of consecutively emitted photons under pulsed p-shell-resonant excitation. The device features a single-photon fiber-coupling efficiency per excitation pulse of up to 53.7(2)\%, corresponding to 1.2\,Million clicks per second at the single-photon detectors at 80\,MHz excitation frequency. Furthermore we demonstrate that the Purcell-enhanced $T_1$ allows for driving the fiber-pigtailed device at 1.28\,GHz excitation clock-rate, providing single indistinguishable photons ($g^{(2)}(0)=0.035(11), V_\mathrm{HOM}=68(7)$\%) at application-relevant GHz clock-rates. 

\section{Device Fabrication}
\label{sec:fabrication}

The QD-hCBG micro-cavity devices used in this work are based on a sample grown by molecular beam epitaxy containing InAs/GaAs QDs emitting between 900-950\,nm at cryogenic temperatures. Using a flip-chip wafer-bonding process, a 170\,nm thick GaAs membrane with embedded QD-layer is hybridly integrated with a dielectric SiO$_2$ layer and a backside gold mirror. For the deterministic integration of pre-selected quantum emitters into numerically optimized hCBG devices (see next section and Supplementary Information~S.I., section S1 for details on the simulations and used structural parameters), we employed a marker-based cathodoluminescence mapping and lithography process. For details on sample growth, processing, and deterministic device fabrication we refer to Ref.~\cite{rickert_high_2024}. To realize a robust high-performance fiber-pigtailed device, the fabricated QD-hCBG cavities are directly and permanently coupled to an ultra-high numerical aperture (UHNA) single-mode fiber (SMF) of the type UHNA3 SMF (fiber core radius $r_\mathrm{core}$=900\,nm, numerical aperture (NA) = 0.35, Coherent Corp.). This specialtiy fiber is spliced (transmission $\sim$95\%~\cite{yin_low_2019, zolnacz_method_2019}) to a standard SM fiber (type 780HP SMF, Coherent Corp.) before the coupling to the device.

Deterministic fiber-pigtailing of individual hCBG devices is achieved by employing the process outlined in the following. In a first step, the fiber glued into a ceramic ferrule is aligned to the target photonic structure with an interferometric positioning procedure: The sample surface is illuminated with a supercontinuum source through this fiber placed at a few hundred nanometer distance from the sample. The interference of light reflected from the sample and partially reflected from the fiber's end face is recorded on a spectrometer. The spectral interference signal is used as feedback to adjust the tilt and height of the fiber with respect to the sample surface plane, and to locate the targeted hCBG structure while scanning the fiber across the sample surface. The precision of fiber-alignment to the targeted hCBG is expected to be below 200\,nm. Next, the fiber facet is placed in physical contact with the sample to ensure stability. To achieve a specific distance between the fiber facet and the hCBG, the fiber-facet is etched prior the coupling procedure within an area of $\sim10\,\mu$m around the fiber-core using a focused Xenon ion beam. In this way, the distance between the target QD-hCBG cavity and the fiber-core can be controlled with a precision of $\pm$50\,nm. In the final fiber-pigtailing step, UV-sensitive adhesive is applied around the fiber-ferrule encapsulating the UHNA3 SMF and cured using a focused UV light source. The FC-device is then transferred to an optical setup for characterization. For details on the fiber splicing,  alignment, and etching as well as investigations on the coupling stability, we refer to Ref.~\cite{zolnacz_method_2019}, applying the methods used in this work to QD mesa structures.

\begin{figure*}[ht]
    \center
	\includegraphics[width=0.80\textwidth]{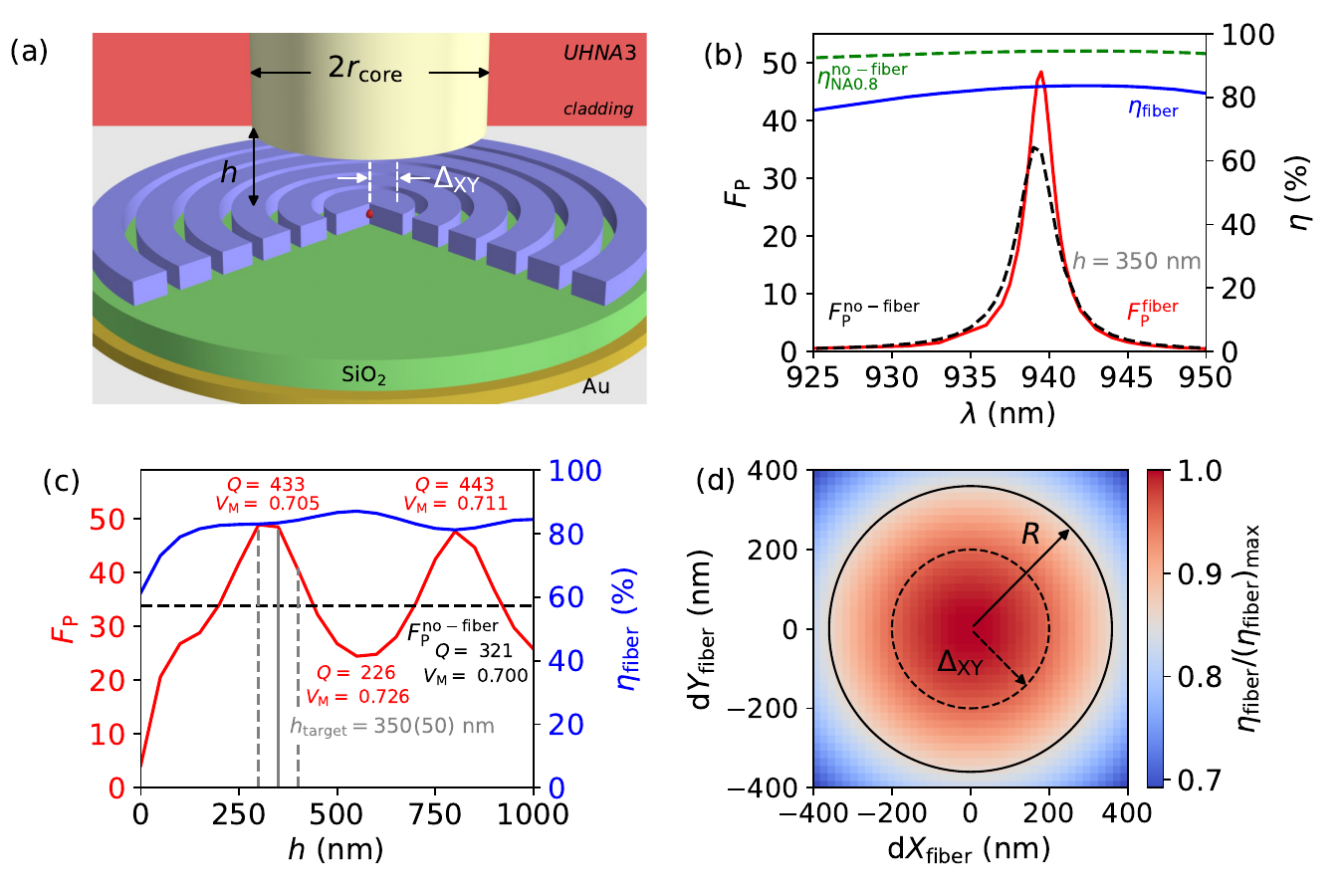}
	\caption{\textbf{Device schematic and simulated performance}. (a) Illustration of the fiber-pigtailed quantum light source with an UHNA3 fiber aligned to the center of the hCBG with embedded QD. The fiber-to-hCBG distance $h$ and the lateral misalignment between CBG and fiber $\Delta_\mathrm{XY}$ are indicated. (b) FEM simulation of Purcell factor $F_\mathrm{P}$ (red) and single-photon fiber-coupling efficiency $\eta_\mathrm{FC\text{-}SPS}$ (blue) of QD emission into the UHNA3 fiber at $h=350$\,nm. The free space performance with $F^\mathrm{no-fiber}_\mathrm{P}$ and lens-efficiency  $\eta^\mathrm{no-fiber}_\mathrm{NA0.8}$ (green) into a lens with NA=0.8 is indicated. (c) Simulated $F_\mathrm{P}$ (red) and $\eta_\mathrm{FC\text{-}SPS}$ (blue) for varying $h$-values at $\lambda$=939.5\,nm. The simulated $F^\mathrm{no-fiber}_\mathrm{P}$ of the QD-hCBG without the fiber is indicated. The target fiber-hCBG distance of $h=350(50)$\,nm is marked. The simulated $Q$ and $V_\mathrm{M}$ for $h$=350, 500, 800\,nm, as well as without fiber are listed. (d) Deviation from the maximum simulated $\eta_\mathrm{FC\text{-}SPS}$ value at $h=350$\,nm and $\lambda$=939.5\,nm for varying lateral displacements $dX_\mathrm{fiber}$/$dY_\mathrm{fiber}$ of the fiber relative to the hCBG cavity's center. The dotted circle denotes a lateral misalignment of $\Delta_\mathrm{XY}=\pm200$\,nm corresponding to the experimentally achieved precision, while the solid line indicates the size of the hCBG's central disc with radius $R$=360\,nm.}
	\label{fig:figure1}
\end{figure*}

\section{FEM Simulations}
\label{sec:Design_and_opti_approach}

Figure~\ref{fig:figure1}(a) shows an illustration of the fiber-pigtailed hCBG device, with the fiber core symmetrically aligned to the hCBG's center, hosting the embedded QD. The fiber is separated by a distance $h$ from the QD-hCBG cavity. For this geometry, the optical performance was calculated using Finite Element Method (FEM) simulations with the commercial software \textit{JCMsuite}~\cite{jcmwave_gmbh_jcmsuite_2024}, using an embedded dipole source to mimic the QD. For details on the FEM simulations, the modelled structure as well as the hCBG cavity design parameters and UHNA3 SMF parameters we refer to the S.I., section~S1.

Fig.~\ref{fig:figure1}(b) shows the simulated optical properties in terms of Purcell factor $F_\mathrm{P}$, free-space performance of the QD-hCBG cavity into a lens without fiber and simulated dipole power coupled to the UHNA3 fiber's two degenerate ground modes. For a target value of $h_\mathrm{target}$ = 350\,nm, simulated $F_\mathrm{P}$-values exceed 30, while direct single photon fiber-coupling efficiencies of $\eta_\mathrm{FC\text{-}SPS} > 80\%$ are reached. The experimentally obtained precision of the distance between the QD-hCBG cavity and the fiber core is $\pm$50 nm [37], which is marked in Fig.~1(c) with dashed lines. This error margin corresponds to the change of the $F_\mathrm{P}$-values in the range of 10-20\% and has a negligible effect on $\eta_\mathrm{FC\text{-}SPS}$. The simulated $\eta_\mathrm{FC\text{-}SPS}$ here corresponds to the coupled power in the two degenerate SMF modes, compared to total emitted power by the embedded dipole, which is equal to the probability of collecting a single photon per excitation pulse, assuming unity QD quantum efficiency and preparation fidelity of the excitation. If Purcell enhancement and efficiency into a lens with NA 0.8 instead of the SMF is considered, the simulated QD-hCBGs performance in absence of the fiber is similar in terms of $F_\mathrm{P}$, with slightly higher efficiencies of $\eta_\mathrm{NA0.8}>93\%$. The lens-efficiency corresponds to the fraction of total dipole power collected in the far-field for the corresponding NA, accordingly.

The fiber-to-CBG distance $h$ has a significant influence on the achievable Purcell enhancement, as can be seen from Fig.~\ref{fig:figure1}(c) for a wavelength of $\lambda=939.5$\,nm. $F_\mathrm{P}$-values between 50 and 25 are observed in the simulations for $200<h<1000$\,nm. This large influence stems from additional vertical confinement to the hCBG mode in vertical direction by the present fiber. As shown in S.I., section~S2, placing the fiber-facet at regions of high (low) hCBG mode intensity in vertical direction, causes a decrease (increase) in the mode's $Q$-factor by up to 45\%. Since the mode volume stays nearly constant, the $h$-dependency of $Q$ leads to an $h$-dependent $F_\mathrm{P}$. The fiber can be considered as effectively becoming part of the hCBG cavity. Noteworthy, the $h$-dependent $Q$ should allow to access the fiber-to-hCBG distance via mode linewidth measurements in the experiment.
Note that the drop in simulated $F_\mathrm{P}$ at h$<$200\,nm originates from a shift in the cavity mode wavelength, since the $h$-dependent simulation is carried out for a constant wavelength. The cavity mode starts to red-shift to longer wavelengths when the fiber comes in close proximity to the hCBG structure, due to increased effective refractive index. $\eta_\mathrm{FC\text{-}SPS}$ is only marginally affected and stays between 80-85\% over the investigated $h$-range, indicating the narrow collimation of the hCBG-mode's emission at these distances.

An additional parameter with a more significant effect on the $\eta_\mathrm{FC\text{-}SPS}$ is the lateral displacement of the UHNA3's fiber-core with respect to the center of the QD-hCBG cavity. The deviation from the maximum efficiency with lateral displacements of the fiber are shown in Fig.~\ref{fig:figure1}(d), with the indicated alignment accuracy of $\pm200$\,nm of the pigtailing process still yielding $>90\%$ of the maximum fiber-efficiency.   

Note that the simulations in Fig.~\ref{fig:figure1} assumed an ideal position of the embedded QD, perfectly centered vertically and laterally in the center hCBG disc. Furthermore, the central mode wavelength $\lambda_\mathrm{C}$ in the hCBG cavity can be varied by $\pm10$\,nm by slightly varying the diameter of the central hCBG disc, without affecting the simulated efficiency significantly. 

\section{Spectral Properties}
\label{sec:performance}

\begin{figure}[t]
    \center
	\includegraphics[width=0.5\textwidth]{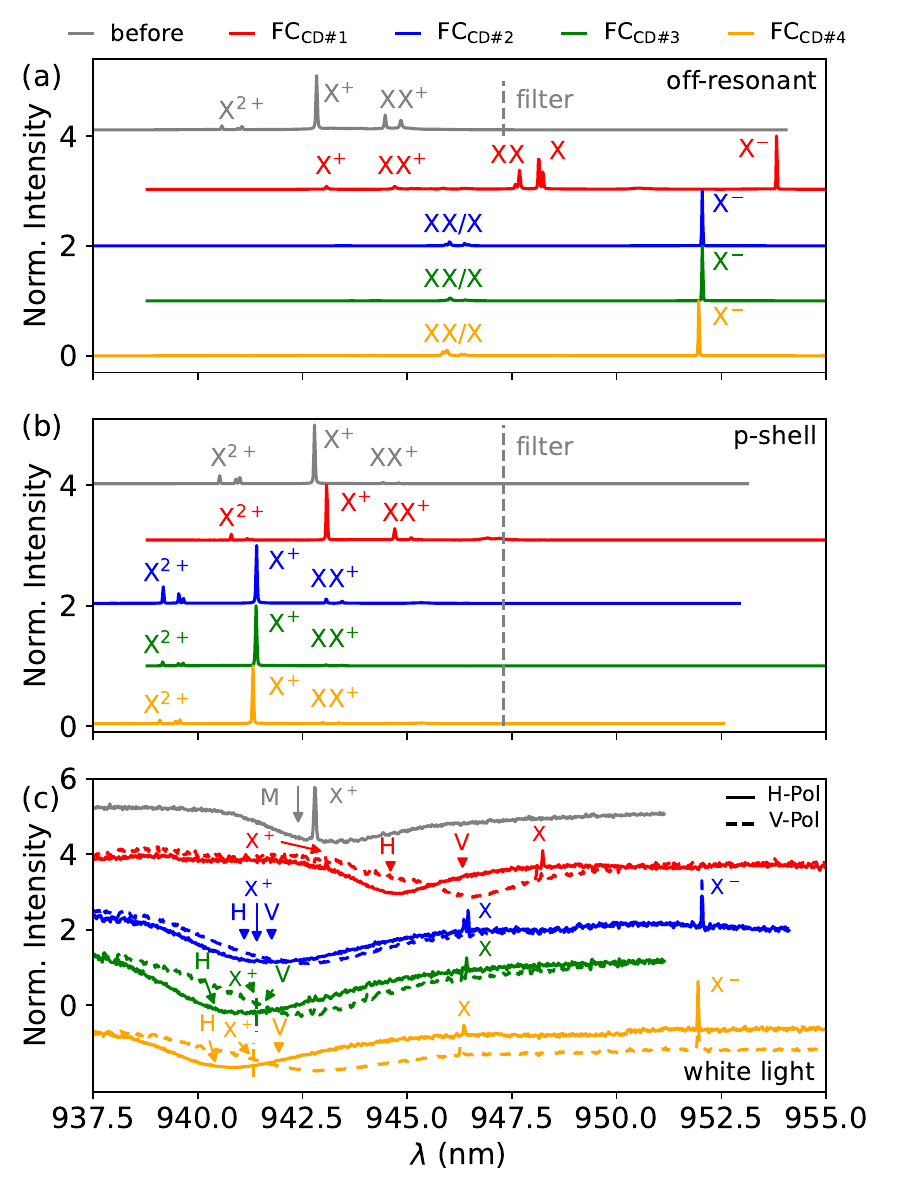}
	\caption{\textbf{Emission properties of the QD-hCBG device before and after fiber-pigtailing.} Shown are measurements from four different cooldowns (FC$_\mathrm{CD}$) \#1 (red), \#2 (blue), \#3 (green) and \#4 (orange) of the fiber-pigtailed device and the hCBG before fiber-coupling as reference (grey). Dashed grey lines indicate the long-wavelength cut-off of an bandpass filter used in the measurements. (a) Normalized emission intensity of the fiber-pigtailed device under pulsed off-resonant excitation ($\lambda_\mathrm{exc}~=~793$\,nm) Respective QD states are indicated. (b) Normalized emission intensity under pulsed p-shell excitation.  (c) Normalized white-light reflection spectra for linear H- (solid) and V-polarisation (dashed line).}
	\label{fig:figure2}
\end{figure}

For quantum-optical experiments, the  fiber-pigtailed QD-hCBG device is transferred into a closed-cycle He-cryostat, and the 780HP side of the 780HP-UHNA3 patchcord is connected to a fiber feedthrough of the cryostat, followed by a controlled cooldown. The room temperature side of the fiber-feedthrough is connected to a 1:2 fiber-coupler (split ratio: 90:10), whose 90\% arm is connected to a fiber-collimator directing the sample emission to a spectrometer. Spectra and time-resolved measurements are recorded using an attached Si-CCD camera or a single photon nanowire detector (efficiency $\sim$85\%), respectively. The 10\% arm of the fiber-coupler serves as input for the excitation laser and is either connected to a $\sim2$\,ps-pulsed tunable excitation laser with 80\,MHz excitation rate, or to a broadband white-light source for reflection measurements. 
The QD-hCBG before the fiber-pigtailing was characterized using the identical setup and excitation sources, but with the sample's emission collected using a NA=0.8 microscope objective in the cryostat via a free-space path towards the spectrometer. Detailed information on the setup can be found in S.I., section~S2.

The fiber-pigtailed QD device was cooled down while excited with off-resonant pulses at excitation wavelength $\lambda_\mathrm{exc}=793\,$nm and the emission was recorded on the spectrometer's Si-CCD camera. Details of the sample's emission during the cooling cycle are shown in {S.I., section~S3, indicating that the fiber-to-hCBG alignment was stable throughout the cooldown, and no plastic deformation occurred. The cryostat's temperature sensor displayed a final temperature of about 4.8\,K, which is 0.8\,K warmer than for the QD-hCBG device without fiber-attached, indicating a potential increased thermal load. However, we confirmed proper thermalization of the device from its emission spectra (see below).

In Figure~\ref{fig:figure2}, spectroscopic data of the selected QD-hCBG cavity before and after the fiber-pigtailing process is presented (sample temperature: $T$=4.8\,K). Fig.~\ref{fig:figure2}(a) depicts spectra of the fiber-pigtailed QD-hCBG emission after four consecutive cooldowns (FC$_\mathrm{CD}$) and the reference spectrum before fiber-pigtailing, each under off-resonant excitation at $\lambda_\mathrm{exc}~=~793$\,nm and excitation power $P_\mathrm{exc}\sim$10-100\,nW. The emission spectrum prior to the pigtailing (grey line) shows dominant positive trion states (multiple lines associated with excitonic X$^{2+}$ and other higher hole states, the dominant X$^+$ and two lines corresponding to biexcitonic XX$^{+}$ transitions), commonly observed for the investigated InAs QDs and dependent on their background doping~\cite{shang_c2v_2020}. The assignment of the emission lines to specific QD states was confirmed by conducting excitation power- and polarization-resolved measurements after the first cooldown of the fiber-pigtailed device (see S.I., section S5. for details). 

The spectrum for the fiber-pigtailed QD-hCBG device under similar excitation conditions of the first cooldown in red, however, appears noticeably different, with dominant QD emission lines identified as the charge-neutral states X and XX as well as the negative trion X$^-$. The observed change to the dominant X$^-$ prevails for the cooldown runs \#2-4, accompanied by a considerable blue-shift relative to cooldown run \#1 and further reduced intensity from the positive trions under off-resonant excitation.

As X$^+$ dominated the spectrum under off-resonant excitation in the uncoupled device, the spectrum was taken with a bandpass filter that focused on  the spectral region around the trion. Therefore the QD states at longer wavelengths are not visible in the reference spectrum before fiber-pigtailing. However, as can be seen in S.I., section~S4, if the intensity scale is logarithmic, it can be seen clearly that the observed emission lines stem from the same QD. 
 
The successful fiber-pigtailing is additionally confirmed by the emission spectra under quasi-resonant excitation in Fig.~\ref{fig:figure2}(b), revealing near-identical spectra throughout cooldown runs \#2-4. Moreover, we observe wavelength shifts for the X$^+$ state of +0.25\,nm for cooldown \#1 and -1.55\,nm for the cooldowns \#2-4 relative to the reference spectrum before fiber-coupling. The wavelength of the excitation laser at $P_\mathrm{exc}$$\sim$3\,$\mu$W was set to the p-shell resonance for the X$^+$ state, as identified from photoluminescence excitation (PLE) scans before and after pigtailing (see  S.I. section~S5). The PLE scans additionally confirm the observed wavelength shifts of the X$^+$ transition by showing a corresponding shift in the $\lambda_\mathrm{exc}$ of the p-shell, indicating that it is in fact a spectral shift resulting from a change of the QD's spectral fingerprint rather than a measurement artifact (e.g. due to changes in the  alignment of the detection path).

Fig.~\ref{fig:figure2}(c) shows white-light reflection measurements before and after pigtailing of the QD-hCBG device. The H- and V-polarized cavity modes are visible as dips in the recorded reflection spectra. In addition, the white-light illumination also excited the QD states, which enables a straightforward analysis of the spectral detuning between cavity and quantum emitter. The spectrum before the pigtailing confirms the good alignment of the X$^+$ emission with the hCBG's cavity mode. For a quantitative analysis, the cavity modes positions are extracted from a Fano fit~\cite{buchinger_optical_2023}, causing the central mode wavelength $\lambda_\mathrm{M}$ to be shifted slightly relative to minima of the reflection dips. Interestingly, the cavity mode appears considerably red-shifted after cooldown run \#1 of the fiber-pigtailed device (compared to the measurements before pigtailing), while a relatively stable blue-shift is observed for the cooldown runs \#2-4.

Due to the fact that both the QD emission and the hCBG modes exhibited a red-shift during cooldown \#1 compared to before the pigtailing, while the cooldowns afterwards showed a blue-shift, a temperature-induced shift is unlikely, not least as the proper thermalization of the fiber-pigtail was confirmed by comparing the temperature dependence of the emission intensity during the cooldown runs with theoretical predictions (see. S.I. section~S3). We thus identify strain as the origin of the observed shifts in emission wavelengths and also the changes in dominant emission lines. A reasonable explanation could be that the induced strain affects the availability of excess holes required for a dominant X$^+$ emission under off-resonant excitation, enabling excess electrons from nearby donors instead causing X$^-$ to be the dominant line. Under p-shell excitation, the X$^+$ transition still remains dominant. The strain is likely built up due to the different thermal expansion coefficient of the hCBG substrate (i.e. GaAs, SiO$_2$ and Au, see Fig.~\ref{fig:figure1}(a)) and fiber-pigtail (i.e. silica and UV-adhesive). The strain could also explain the shifts in mode wavelengths between the cooldowns, by changing the (effective) refractive index of the combined fiber-hCBG microcavity. For a rough estimation based on FEM simulations, we find that a refractive index change $\Delta{n}<$0.2\% is already sufficient to create the observed mode shifts. Such a $\Delta{n}$ is well within reach for already small amounts of strain~\cite{tran_systematic_2016}.

\begin{figure*}[ht]
    \center
	\includegraphics[width=0.85\textwidth]{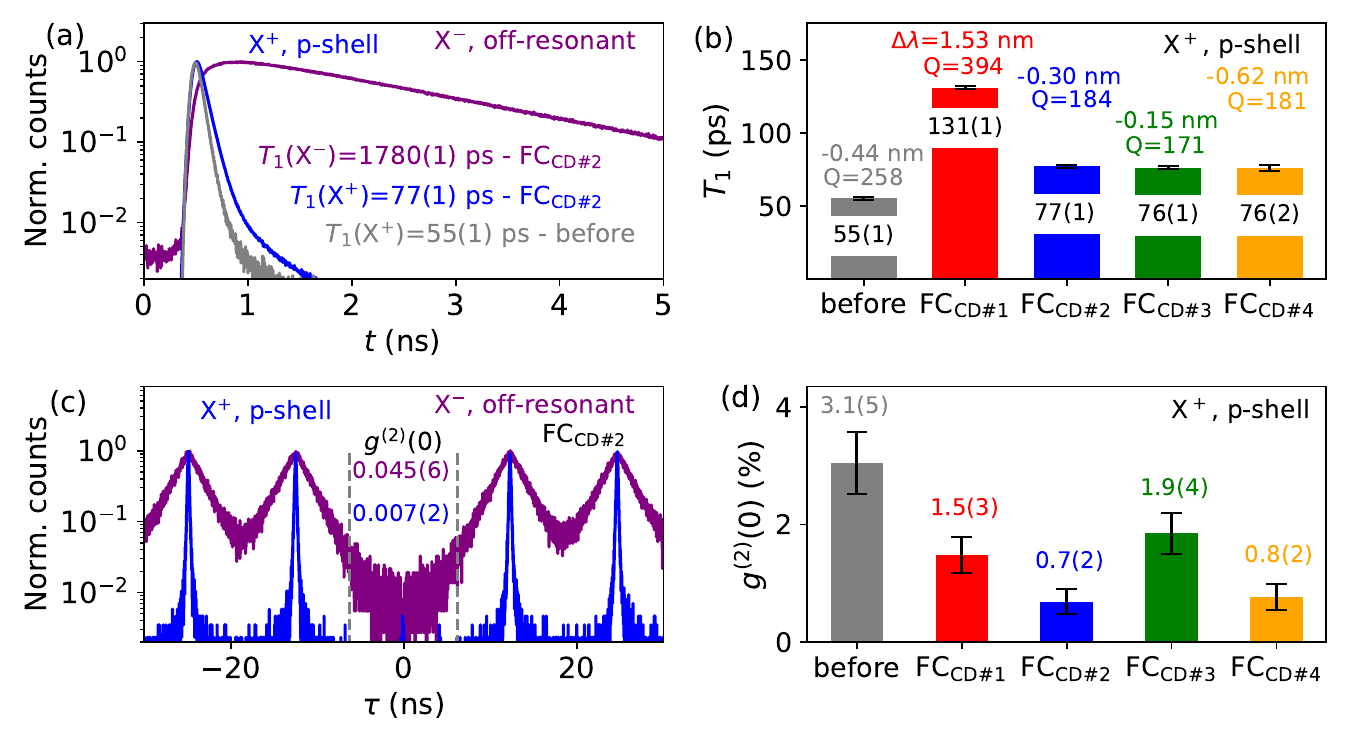}
	\caption{\textbf{Time-resolved measurements before and after the pigtailing.} (a) Lifetime-measurement with extracted $T_1$-times from exponential fits for X$^+$ in p-shell excitation before (grey) and after (blue) the fiber-pigtailing. The measurement for the X$^-$ after the pigtailing (purple) in off-resonant excitation is also shown. (b) Measured $T_1$-times of X$^+$, mode $Q$-factors and QD-mode spectral mismatch $\Delta\lambda$. (b) Second-order-auto-correlation $g^{(2)}(\tau)$-measurements and $g^{(2)}(0)$-values from comparing the integrated events at $\tau=0$ to the integrated neighbouring peaks at 12.5~ns time-window. Measurements for X$^+$ and X$^-$ for respective excitation conditions are shown. The integration time-window around $\tau=0$~ns is indicated. (d) $g^{(2)}(0)$-values of the X$^+$ under p-shell excitation before the fiber-pigtailing and for pigtailed cooldowns.}
	\label{fig:figure3}
\end{figure*}

Strain-induced changes of the emission wavelength of InAs QDs have been reported for monolithic samples in combination with permanently adhered fiber-systems~\cite{shang_single-_2022}. The significant change from predominantly positive to negative trion emission observed in our work for the hybrid samples under off-resonant excitation, however, was not reported yet. While a detailed understanding of these changes in state-occupation will require further research efforts, the fact that X$^+$ and X$^-$ show these variations might point towards background dopants or defect levels providing excess-charge-carriers close to the QD being affected by the strain. Furthermore, the X$^+$ emission being inhibited under off-resonant excitation at $\lambda_\mathrm{exc}$=793\,nm, while showing considerably brighter emission in p-shell excitation, hints towards the specific energy levels of these defects.
A possible factor that benefits the observation of these effects in the present hCBG sample, might be its thin GaAs-membrane on thin SiO$_2$ spacer, enabling an efficient strain-transfer from the glued fiber to the QD position. Since the central disc is free-standing and should be neither in contact with glue nor fiber (due to the ion-beam milling of the fiber core and cladding), this strain-transfer is supported by the underlying layers which extend over a larger areas covered by ferrule and glue.

It is worth noting that both the X$^+$ and mode wavelengths appear to remain constant after the first cooling and warming cycle. Especially the cavity-mode wavelengths after the  cooldown runs \#2-4 of the fiber-pigtailed device are more similar to each other, indicating that the strain conditions are still different compared to before, but remain constant after the first temperature cycle.

\section{Efficiency and Purcell Enhancement}

After confirming that the employed fiber-pigtailing method allows for deterministic coupling of a selected QD-hCBG, we now investigate the quantum optical properties of the fiber-pigtailed QD device.
Fig~\ref{fig:figure3}(a) shows time-resolved measurements during cooldown~\#2 for the X$^+$ under p-shell excitation, and for X$^-$ under off-resonant excitation. Additionally, a time-resolved measurement of the X$^+$ before the pigtailing is shown. A $T_1(\mathrm{X}^+)$=77(1)\,ps for the FC-device under p-shell excitation is extracted by an exponential fit. This is slightly longer than the 55(1)~ps before the pigtailing. 
A typical $T_1$ time of positive trions in p-shell excitation in the surrounding membrane of the sample is 680(111)\,ps, which corresponds to $F_\mathrm{P}=12.4(2.2)$ before, and $8.8(1.4)$ for the fiber-pigtailed device. $T_1$(X$^-$) under off-resonant excitation is found to be 1.780(1)\,ns, which is attributed to a large spectral mismatch of X$^-$ to the mode (see. Fig,~\ref{fig:figure2}(c)).

The measured $T_1$-times of the X$^+$ for the consecutive cooldowns of the pigtailed device are displayed in Fig.~\ref{fig:figure3}(d). The observed spectral mismatch of X$^+$ wavelength and closest cavity mode and the measured cavity mode $Q$-factor as $\lambda_\mathrm{M}/w_\mathrm{M}$ are additionally listed. The pigtailed device shows a considerably larger $T_1$(X$^+$) during cooldown \#1, while interestingly also exhibiting a clearly increased $Q$ compared to before the pigtailing. However, while the spectral mismatch between X$^+$ and mode was -0.44(0.50)\,nm before, the pigtailed device showed a mismatch of +1.53(0.50)\,nm during the first cooldown. We attribute the longer $T_1$-time to the larger spectral mismatch after the first cooldown.
With the strain conditions settled after the first temperature cycle, the following cooldowns yield very similar $T_1$ times, alongside X$^+$-mode mismatches between -0.62(0.50)\,nm and 0.15(0.50)\,nm. While the observed spectral mismatch is close to the value before the pigtailing, the $Q$-factors are clearly reduced for the cooldowns \#2 to \#4. We attribute the slightly higher $T_1$ times after the fiber-pigtailing to the decreased Purcell enhancement by the reduced $Q$ caused by the distance-dependent effect of the fiber on the $F_\mathrm{P}$ here as predicted in Fig.~1(\st{d}\textcolor{red}{c}).

The decrease in experimental $Q$ between before the pigtailing and for cooldown \#1 is approximately $390/260\simeq1.35$, which is also close to the relation between the simulated $Q$ before the pigtailing and at $h_\mathrm{target}=350$\,nm of 430/320, considering that the overall $Q$ was reduced slightly through imperfect fabrication. This indicates that before cooldown \#1, the distance of fiber and hCBG cavity was potentially close to the target value. After the warmup and following cooldowns \#2 to \#4, the experimental $Q$ drops to 0.71$\times$ the $Q$-factor before the pigtailing, which means that the fiber-to-hCBG distance for the pitailed device afterwards might be up to 200\,nm larger than the $h_\mathrm{target}$ of 350\,nm. 
We note that the expected strain-related refractive index change causing the spectral shift of the modes results in a negligible change in $Q$ as confirmed from the simulations, hinting further towards a fiber-distance related effect.
The resulting $T_1$-times before and after the pigtailing during cooldowns \#2 to \#4 exhibit a similar relation of 0.72, implying that the change in $T_1$ could stem from the distance-dependent reduced $F_\mathrm{P}$ by the lower $Q$. Future work simulating the strain conditions~\cite{schliwa_impact_2007} could answer the question of how much $T_1$ is additionally affected by the strain itself. Note that the device does not reach the simulated $F_\mathrm{P}$ of 30 and more already before the coupling, likely due to a non-ideal spatial integration of the QD. 

To proof the non-classical photon statistics of the photons emitted by the fiber-pigtailed QD device, we performed experiments in Hanbury-Brown-and-Twiss (HBT) type configuration. Fig.~\ref{fig:figure3}(c) shows the resulting second-order auto-correlation histograms $g^{(2)}(\tau)$ and corresponding $g^{(2)}(0)$-values obtained by integrating the coincidences within a 12.5\,ns wide window around $\tau~=~0$ and comparing this number with the integrated coincidences of the neighbouring peaks.

A multiphoton-suppression value of $g^{(2)}(0)~=~0.007(2)$ is obtained for the X$^+$ under p-shell excitation, while the X$^-$ shows $g^{(2)}(0)~=~0.045(6)$ in off-resonant excitation. The higher multiphoton suppression obtained for X$^+$ is benefited by the possible quasi-resonant excitation compared to the above-band excitation for X$^-$, and also the significantly shorter $T_1$(X$^+$). For X$^-$, the long $T_1$ time leads to significant residual events around $\tau$=0 in the $g^{(2)}(\tau)$-measurement at 12.5\,ns pulse separation.

As a final comparison of before and after the pigtailing, Fig.~\ref{fig:figure3}(d) shows the obtained $g^{(2)}(0)$-values pf X$^+$ for each case. The multi-photon suppression in the fiber-pigtailed cases are slightly better than before the pigtailing, with comparably small error due to the short lifetime and thus high statistics in the corresponding time-bins of the histograms. The difference between the before and pigtailed case is most likely not significant, given the fact that the measurements were some time apart, leading to different lab conditions like stray light, and detector dark counts. The error for the respective $g^{(2)}(0)$-values does only account for the statistical error for the integration of the histogram, and does not take such systematic deviations into account.

The measured multiphoton suppression for the pigtailed QD-hCBG device of below 1\%, obtained via direct integration of the measured $g^{(2)}(\tau)$-histogram, is the best performance for a directly fiber-coupled device under these excitation conditions reported so far, if compared to the value of $\sim$1.5\% in Ref. \cite{northeast_optical_2021} by Northeast \textit{et. al.} and $\sim$3.7\% in Ref.~\cite{snijders_fiber-coupled_2018} by Snijders \textit{et. al.}. It should be noted, however, that we used a slightly narrower spectral filter (bandwidth: 100\,$\mu$eV or 0.07\,nm) in our experiments compared to Ref.~\cite{northeast_optical_2021} ($\sim$0.1\,nm), while the experiments in Ref.~\cite{snijders_fiber-coupled_2018} did not use any spectral filtering at all, which was possible thanks to the resonance fluorescence excitation scheme (cross-polarized excitation-detection configuration) employed in their fiber-pigtailed device. Since arbitrarily high spectral filtering will always improve the measured $g^{(2)}$-value, we point out that the 100\,$\mu$eV filtering settings are still 2-3 times broader than the zero phonon line of the pigtailed X$^+$, and only discard small portions of the phonon sideband, a filtering comparable~\cite{liu_single_2018} or less strict~\cite{somaschi_near-optimal_2016,wang_-demand_2019} than other works in literature.

After confirming the single-photon nature of the fiber-pigtailed QD device's emission, its efficiency can be estimated by analyzing the photon flux detected at the single photon detector in the experimental setup and taking into account transmission losses from the fiberpigtail to the detection system. The observed countrates $R^\mathrm{before}$ before and $R^\mathrm{FC}$ after the pigtailing (in this case for cooldown \#2) are listed in Table~\ref{tab:countrates}. A countrate of $R^\mathrm{FC}_{\mathrm{X}^+}=300(10)$\,kcps (kilo clicks per second) is obtained for the X$^+$ under p-shell excitation ($P_\mathrm{Sat}\sim3\,\mu$W) on the SNSPDs, while the X$^-$ reaches $R^\mathrm{FC}_{\mathrm{X}^-}=1.20(5)$\,Mcps at off-resonant excitation with $\lambda_\mathrm{exc}=793$\,nm ($P_\mathrm{Sat}<15\,$nW).

\begin{table}[ht]
\begin{center}
\caption{Measured countrates $R$ and intensities $I$ in Megaclicks/second (Mcps) before and after the pigtailing for cooldown \#2 for given QD states and excitation on CCD and SNSPDs.}
\label{tab:countrates}
\begin{tabular}{||c c c||}
\hline
 & X$^+$ (p-shell) & X$^-$ (off-resonant)\\
\hline \hline
$R^\mathrm{FC}$(SNSPD) & 0.30(1)\,Mcps & 1.2(1)\,Mcps \\
\hline
$I^\mathrm{FC}$(CCD) & 0.175(5)\,Mcps & 0.450(5)\,Mcps \\
\hline
$I^\mathrm{before}$(CCD) & 0.252(5)\,Mcps & - \\
\hline
\end{tabular}
\end{center}
\end{table}

By dividing the measured countrates from Table~\ref{tab:countrates} by the excitation repetition rate of $f=80$\,MHz, the end-to-end efficiency at the detectors is $\eta_\mathrm{overall}=R/f$, yielding $\eta_\mathrm{overall}(\mathrm{X}^+)~=~0.375$\% and  $\eta_\mathrm{overall}(\mathrm{X}^-)~=~1.50$\%.
To compare these efficiencies to the simulated $\eta_\mathrm{FC\text{-}SPS}$ in Fig.~\ref{fig:figure1}(a), we take the setup efficiency $\eta_\mathrm{setup}$ into account. Details about the efficiency estimation can be found in S.I., section~S8. Considering the UHNA3-to-780HP splice $\eta_\mathrm{splice}=0.95(1)$, the fiber-feedthrough at the cryostat $\eta_\mathrm{cryo}=0.60(5)$ and transmission losses by the fiber-beamsplitter, several fiber-connectors, the spectrometer and the connection to the SNSPDs $\eta_\mathrm{detection}~=~0.049(1)$, we obtain an experimental $\eta_\mathrm{FC\text{-}SPS}$ into the pigtailed UHNA3 fiber of$\eta^{\mathrm{X}^+}_\mathrm{fiber}=13.4(2)$\% and $\eta^{\mathrm{X}^-}_\mathrm{fiber}=53.7(2)$\% for $\mathrm{X}^+$ and $\mathrm{X}^-$, respectively.

The observed $\eta_\mathrm{FC\text{-}SPS}$ for X$^-$ is already surprisingly close to the simulated efficiency of $\eta_\mathrm{FC\text{-}SPS}=85$\%, a possible non-ideal lateral alignment between fiber and hCBG device. Furthermore, the off-resonant excitation of the X$^-$ reduces the measured efficiency, since e.g. the neutral states are also excited (see cooldown \#2 in Fig.~\ref{fig:figure2}(a)).
Note here, that the high efficiency observed for the X$^-$ state besides it's far detuning from the cavity modes highlights the potential of the broadband capabilities of hCBG devices also for applications benefiting from long radiative lifetimes. The large spectral cavity-emitter detuning substantially increases $T_1$(X$^-$), while $\eta_\mathrm{FC\text{-}SPS}$ remains on a high level.

The observed textcolor{red}{$\eta_\mathrm{FC\text{-}SPS}$} of the X$^+$ is significantly lower than expected, however this can be explained considering the fact, that the X$^+$ was already limited in brightness prior to the pigtailing: Table~\ref{tab:countrates} shows the observed peak-intensities at the CCD for the X$^+$ under p-shell near saturation before the pigtailing with $I^\mathrm{before}_{\mathrm{X}^+}=252(10)$\,kcps, and with pigtailed fiber as $I^\mathrm{FC}_{\mathrm{X}^+}=175(10)$\,kcps. Free-space and fiber-setup efficiencies to the CCD were comparable for the measurements, so that the ratio of peak-CCD counts for the X$^+$ before and after the pigtailing can also act as a rough estimate for $\eta_\mathrm{FC\text{-}SPS}$. $I^\mathrm{FC}_{\mathrm{X}^+}/I^\mathrm{before}_{\mathrm{X}^+}$ yields $\sim0.69(4)$, which is reasonably close to the simulated $\eta_\mathrm{FC\text{-}SPS}/\eta^\mathrm{nofiber}_\mathrm{NA0.8}=0.91$ bearing the aforementioned experimental limitation in mind and thus further indicating a high degree of pigtailing precision. We observed, that the saturated X$^+$ brightness for the incorporated QDs in the hCBGs before the pigtailing varied by up to a factor of 4 between QDs, which we attribute to changing dopant and defect layer environments, that provide the charge carries for the X$^+$. The fiber-pigtailed device here was primarily chosen for the high Purcell enhancement of the X$^+$, rather than the X$^+$ brightness, and we plan to optimize both parameters in the future.

The performance of the fiber-pigtailed QD device presented above compares favorably with an earlier report on CBG-based plug-and-play sources by Jeon et al.~\cite{jeon_plugandplay_2022}. We achieve significantly higher $F_\mathrm{P}$ and lower $g^{(2)}(0)$ for the quasi-resonantly excited X$^+$. The improvement in multiphoton-suppression we thereby attribute to the quasi-resonant excitation scheme implemented in our work. Moreover, our fully deterministic technology for QD-device integration and fiber-pigtailing enables us to increase $\eta_\mathrm{FC\text{-}SPS}$ of the X$^+$ and especially X$^-$ state by more than a factor six. This advancement is further facilitated by the optimized hybrid back-reflector design in combination with the UHNA3-fiber and precise fiber-to-hCBG alignment accuracy.

\begin{figure}[ht]
    \center \includegraphics[width=0.5\textwidth]{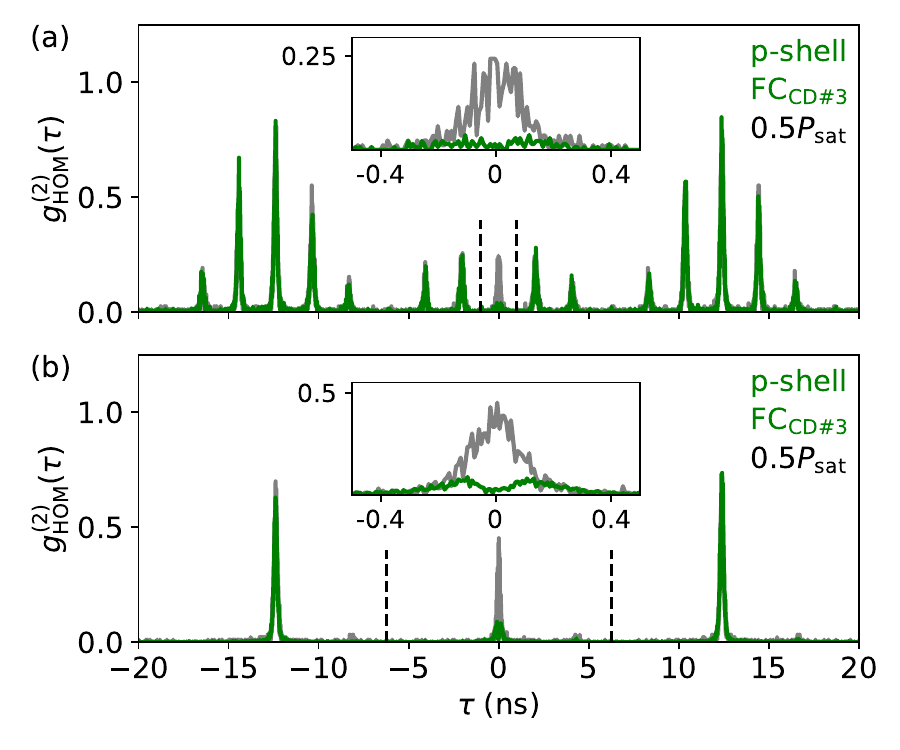}
	\caption{\textbf{Two-photon interference measurements for the FC-device.} The $g^{(2)}_\mathrm{HOM}(\tau)$ measurements were taken during FC$_\mathrm{CD\#3}$ under 80\,MHz p-shell excitation at half the saturation power $P_\mathrm{sat}$. Co-polarized measurements are shown in green, cross-polarized measurements in grey. (a) for a separation of $\delta{t}=2$\,ns for the exciting pulses. (b) for a separation of $\delta{t}=12.5$\,ns for the exciting pulses. The HOM visibility $V_\mathrm{HOM}$ is obtained by comparing the co- and cross polarized peak areas at $\tau=0$. The respective integration time-windows are indicated with dashed lines.}
	\label{fig:figure4}
\end{figure}

\section{Photon-Indistinguishability}
\label{sec:indistinguishability}

Next, we investigated the photon-indistinguishability of the emission of the fiber-pigtailed device using two-photon interference experiments to access $g^{(2)}_\mathrm{HOM}(\tau)$ in a Hong-Ou-Mandel (HOM) setup. For this purpose we interfered consecutively emitted X$^+$ photons in the HOM-setup after spectral filtering via the spectrometer (see {S.I., section S2 for experimental details). To quantify the degree of indistinguishability, measurements in co- and cross-polarized configuration are compared. The HOM-experiments are conducted during cooldown \#3 for two different temporal delays $\delta{t}$ of 2\,ns and 12.5\,ns for consecutively emitted photons, providing additional insights in possible dephasing mechanisms \cite{rickert_high-performance_2023}. 

The resulting HOM-histograms are shown in Figure~\ref{fig:figure4}, while the extracted two-photon visibilities $V_\mathrm{HOM}$ are summarized in Table~\ref{tab:VHOM_80MHz}. Figure~\ref{fig:figure4}(a) yields $V_\mathrm{HOM}~=~0.78(4)$ for p-shell excitation at $0.5P_\mathrm{sat}$ and $\delta{t}=2$\,ns. Accounting for residual multi-photon emission events, we obtain a corrected value of $V^\mathrm{corr}_\mathrm{HOM}=(1~+~2g^{(2)}(0))V_\mathrm{HOM}=0.82(4)$~\cite{zhai_quantum_2022}, using the $g^{(2)}(0)$-value measured during cooldown run~\#3 (c.f., Fig.~\ref{fig:figure3}(d)). Fig.~\ref{fig:figure4}(b) shows $g^{(2)}_\mathrm{HOM}$ under the similar conditions, but for $\delta{t}=12.5$\,ns. The extracted visibility is slightly reduced to $V_\mathrm{HOM} = 0.75(4)$ ($V^\mathrm{corr}_\mathrm{HOM}~=~0.79(4)$) at these increased time delays, a commonly observed phenomenon~\cite{thoma_exploring_2016} that could be reduced in future works by employing coherent excitation schemes~\cite{wang_near-transform-limited_2016}. At saturation power, values of $V_\mathrm{HOM} = 0.63(4)$ ($V^\mathrm{corr}_\mathrm{HOM}~=~0.67(4)$) are measured for $\delta{t}=2$\,ns and $V_\mathrm{HOM}~=~0.52(4)$ ($V^\mathrm{corr}_\mathrm{HOM}=0.55(4)$) for $\delta{t}=12.5$\,ns. The further decrease in indistinguishability at elevated excitation power indicates increased decoherence due to excitation power induced dephasing. We note again, that the spectral filtering applied in the measurements conducted here partially filtered out the phonon side bands of the X$^+$ emission, but collected all emission from the zero phonon line.

\begin{table}
\begin{center}
\begin{tabular}{||c c c c ||}
\hline
$P/P_\mathrm{sat}$ & $\delta{t}$ (ns) & $V_\mathrm{HOM}$ & $V^\mathrm{corr}_\mathrm{HOM}$\\
\hline \hline
\multirow{2}{*}{0.5} & 2 & 0.78(4) & 0.82(4) \\
\cline{2-4}
 & 12.5 & 0.75(4) & 0.79(4) \\
\hline

\hline
\multirow{2}{*}{1} & 2 & 0.63(4) & 0.67(4) \\
\cline{2-4}
 & 12.5 & 0.52(4) & 0.55(4) \\
\hline

\end{tabular}
\end{center}
\caption{Two-photon HOM visibilities $V_\mathrm{HOM}$ of the FC-device during cooldown run~\#2 under p-shell excitation at given fraction of saturation Power $P_\mathrm{sat}$ and excitation and detection time delay $\delta{t}$.}
\label{tab:VHOM_80MHz}
\end{table}

Although experimental data which would allow for a direct comparison of the photon-indistinguishability before the fiber-pigtailing is not available, we can still compare the performance of the pigtailed device with  experimental data of free-space coupled devices under comparable excitation conditions stemming from the same wafer (see S.I., section~S7}. The HOM-results summarized in Table~\ref{tab:VHOM_80MHz} yield very similar $V_\mathrm{HOM}$ for given $T_1$-times and inhomogeneous broadening under p-shell excitation, indicating that the photon-indistinguishability is mostly affected by the effects of the fiber-pigtailing on the emitter's $T_1$-time as well as the degree of inhomogeneous broadening.

\section{GHz Clock-rate Operation}
\label{sec:GHz}

Furthermore, we demonstrate that the short Purcell-enhanced radiative lifetime of the fiber-pigtailed QD-hCBG cavity enables its operation at clock-rates in the GHz-regime. Using a home-built frequency multiplication setup, we reduced the original repetition period of the 80~MHz laser system from $\delta{t}=12.5$\,ns down to $\delta{t}=781$\,ps corresponding to an excitation repetition rate of $f=1.28$\,GHz. 

\begin{figure}[ht]
    \center
	\includegraphics[width=0.5\textwidth]{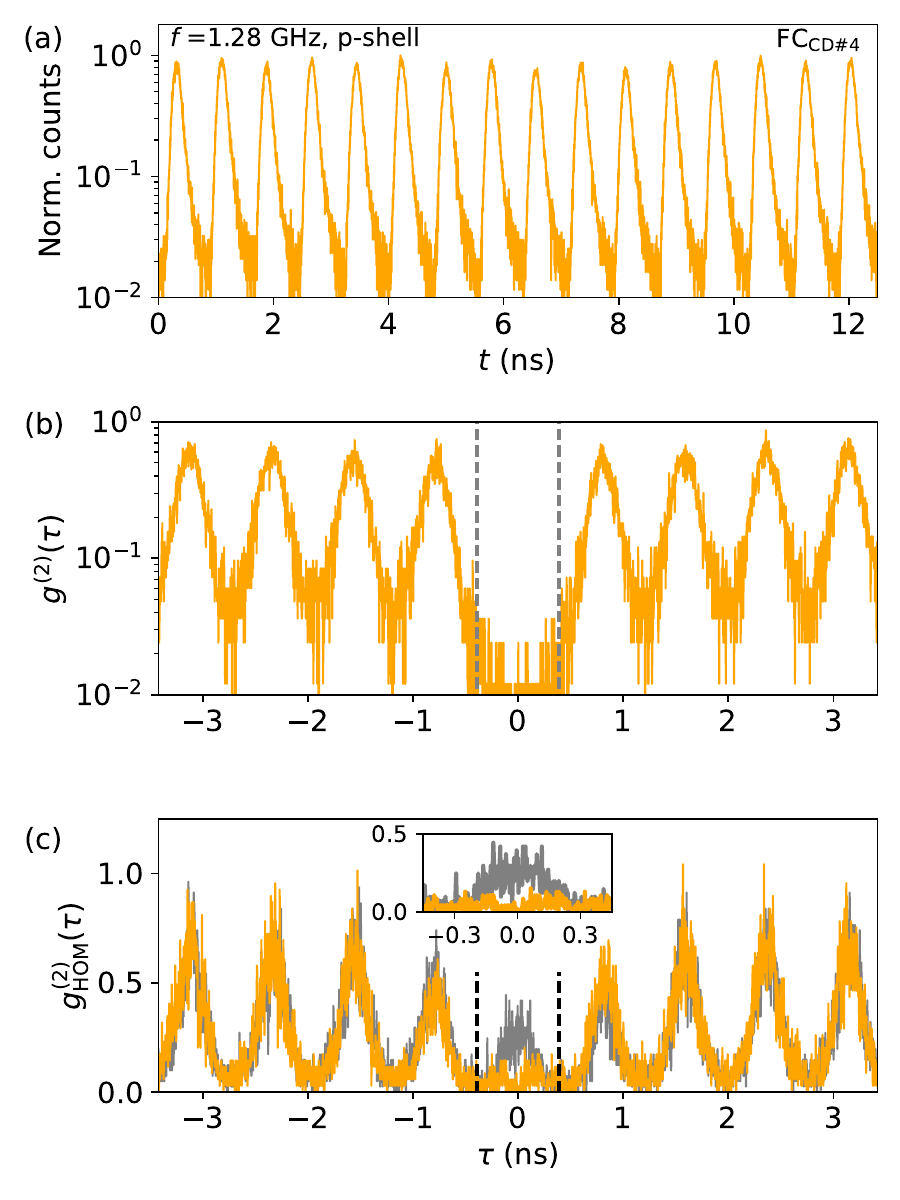}
	\caption{\textbf{Performance of the fiber-pigtailed QD device (X$^+$ emission) under p-shell excitation at a clock-rate of $f$=1.28\,GHz.} The temporal delay between consecutive single-photon amounts to $1/f=781$\,ps. (a) Time-resolved trace of a single-photon pulse train in logarithmic scaling. (b) Photon-autocorrelation $g^{(2)}(\tau)$-measurement. (c) Two-photon interference $g^{(2)}_\mathrm{HOM}(t)$ histograms measured in co- (orange) and cross- (grey) polarized configuration at $P_\mathrm{exc}$=$P_\mathrm{sat}/16$. Dashed lines indicate the $781$\,ps repetition period.}
	\label{fig:figure5}
\end{figure}

Time-resolved measurements of the fiber-pigtailed QD device under p-shell excitation at these GHz frequencies during cooldown run~\#4 are displayed in Figure~\ref{fig:figure5}(a). The short $T_1$ time of the X$^+$ transition enables a clear separation between consecutive single-photon pulses even at 1.28\,GHz clock-rate. Note, that the X$^+$-state has been operated far below saturation in this measurement ($P_\mathrm{sat}/16$), due to technical limitations for the transmission of the used self-built frequency multiplication setup. 

The corresponding $g^{(2)}(\tau)$-measurement is depicted in Fig.~\ref{fig:figure5}(b), with an extracted $g^{(2)}(0)$-value of 0.035(11), obtained by direct coincidence integration and comparison to neighbouring peaks in a temporal window of $1/f=\delta{t}=0.781$\,ns. The multiphoton suppression is slightly elevated compared to operation at 80\,MHz (compare Fig.~\ref{fig:figure3}(d)). This is due to the slower bi-exponential decay component of $>600$\,ps in the time-trace for the X$^+$ under p-shell excitation, resulting in a small but noticeable overlap of consecutive single-photon pulses.

Finally, we also conducted two-photon interference experiments at 1.28\,GHz clock-rate. The HOM-results are displayed in Fig.~\ref{fig:figure5}(c) for co- and cross-polarized measurement configurations. From the raw experimental data we extract an photon-indistinguishability of $V_\mathrm{HOM} = 0.61(7)$) at $\delta{t}=0.781$\,ns and accounting for the finite $g^{(2)}(0)$ yields a corrected value of $V^{corr}_\mathrm{HOM} = 0.68(7)$).

The obtained values are similar to the photon-indistinguishabilities obtained under 80\,MHz p-shell excitation and at significantly higher excitation powers. Considering that both the excitation power per pulse as well as the temporal-separation between interfering photons is considerably smaller, the two-photon interference visibility observed under GHz-driving is lower than intuitively expected.

The aforementioned longer decay component leads to some degree of overlap of neighbouring pulses, reducing the measured indistinguishability. In general, the effect of GHz-driving on the indistinguishability of emitted photons by a quantum emitter is far less explored compared to conventional 80\,MHz excitation rates. Recent results indicate that the short time-scales between consecutive pulses can affect the emission from charged QD states~\cite{rickert_high_2024}, which might well be connected to the limited two-photon interference visibility under GHz-drive observed above. 

\section{Discussion}
\label{sec:discussion}

In summary, we reported on the fully deterministic fabrication of a directly fiber-pigtailed Purcell-enhanced QD device based on a numerically optimized hCBG microcavity coupled to an UHNA3 single-mode fiber and demonstrate its capability for producing single indistinguishable photons at GHz clock-rates. The achieved Purcell factor of $\sim 9$ results in short radiative emission lifetimes $<80$\,ps and a strong multiphoton suppression reflected in $g^{(2)}(0)<1\%$. The fiber-pigtailed device exhibits photon-indistinguishabilities of 55-80\% at 80\,MHz excitation repetition rate under quasi-resonant excitation and we demonstrate a single-photon fiber-coupling efficiency $>53\%$. Moreover, the significant Purcell enhancement enables operation of the fiber-pigtailed device at an excitation clock-rate of 1.28\,GHz, resulting in antibunching values $<4\%$ and photon-indistinguishabilities $>67\%$ under quasi-resonant p-shell excitation.
The results presented in this work clearly demonstrate that cutting-edge QD devices with excellent quantum-optical performance can be fully-deterministically integrated with optical SM fibers for the development of robust and practical quantum-light sources. Integrating the fiber-pigtailed device in compact and user-friendly cryocoolers in the future, will enable the implementation of high-performance quantum light source in field-applications of quantum information science.

To further improve the performance of this type of fiber-pigtailed device, excitation schemes allowing for the coherent pumping of the embedded QD are beneficial. While resonant excitation can be used to produce photons with photon-indistinguishabilities near unity, its realization in all-fiber coupled scenarios remains challenging, as cross-polarized excitation-detection is required with high extinction ratios - as task difficult to achieve in optical fibers. This makes phonon assisted~\cite{quilter_phonon-assisted_2015,ardelt_dissipative_2014} excitation or the recently proposed SUPER scheme~\cite{bracht_swing-up_2021,boos_coherent_2024} good candidates to further push the performance of fiber-pigtailed devices in this context. 
If the emission of the neutral exciton is collected rather than the positive trion emission, stimulated two-photon resonant excitation~\cite{yan_double-pulse_2022,sbresny_stimulated_2022,wei_tailoring_2022} could be used to achieve coherent excitation with all-fiber compatible spectral filtering conditions. 

Another route to improve the degree of coherence under quasi-resonant excitation is a further reduction of the fiber-pigtailed QD-hCBG microcavity system's $T_1$-time. While the $T_1<80\,$ps observed in this work for the X$^+$ transition was limited already before the fiber-pigtailing procedure by non-ideal QD-positioning in the hCBG center disc, optimal positioning of the quantum emitter enables $F_\mathrm{P}>$25 and $T_1<$30\,ps as recently reported in Ref.~\cite{rickert_high_2024}. Modified optimized device designs can additionally increase the Purcell enhancement by a factor of 2-3~\cite{rickert_optimized_2019}, while further enhancement might be possible considering the influence of the optical fiber on the cavities $Q$-factor, if the fiber-to-hCBG distance can be controlled more precisely.   
In addition, the further reduced $T_1$-time would also enhance the temperature-resilience of the photon-indistinguishability enabling the generation of indistinguishable single photons at temperatures achievable with ultra-compact mechanical cryocoolers.

Concerning the obtained $\eta_\mathrm{FC\text{-}SPS}$, the observed strain-attributed wavelength shifts and influences on the emission lines limited the countrates achieved with the fast X$^+$-transition in the current device, but the Mcps countrate of the X$^-$ doubtlessly demonstrated the potential of fiber-pigtailed hCBG cavities for highest coupling efficiencies. To give insight on the repeatability of the coupling technique and the influence of strain, we present experimental data of a second fiber-pigtailed device in  S.I., section S10. Also this second device allowed for unambiguous coupling of the respective target QD-hCBG, but showed a plastic deformation of the fiber-to-cavity alignment during its first cooldown, limiting the efficiency. However, the plastic deformation indicates strain relaxation, and in fact this sample did not show the unexpected emission intensity shifts in off-resonant excitation. In this context, additional investigations are required, for example by minimizing the amount of adhesive, or the membrane area in contact. Alternatively, the strain could be passively countered by an additional adhesion layer on the sample backside, which would exert competing strain~\cite{shang_single-_2022}. Furthermore, the strain could also be actively controlled by fabricating the device bonded to a piezo substrate~\cite{moczala-dusanowska_strain-tunable_2020,rota_source_2024}. Moreover, it is worth mentioning that the theoretical $\eta_\mathrm{FC\text{-}SPS}$ of ~85\% is currently limited by the UHNA3 fiber, which is not optimized for the wavelength range around ~930-950\,nm just above the cut-off wavelength. Moving to telecom O- and C-Band wavelengths will further increase the hCBGs mode overlap to the UHNA3 optical mode-field, therefore boosting the fiber-coupling efficiency to above 90\%~\cite{ma_circular_2024}. The shorter wavelength designs would benefit from fibers that have even smaller core diameters.

Furthermore, future work will also aim for increasing the device functionality of fiber-pigtailed hCBG microcavities. Implementing electrical control via gates, for example, is an interesting route, especially as the presented fully deterministic fabrication- and fiber-pigtailing process is straight-forwardly compatible with schemes enabling spectrally tunable emitter wavelengths~\cite{wijitpatima_bright_2024} and high Purcell enhancement~\cite{barbiero_design_2022,buchinger_optical_2023,rickert_high-performance_2023,ma_circular_2024}.

Finally, the first experiments on GHz clocking reported in this work for a Purcell-enhanced fiber-pigtailed QD device opens the door for high-performance implementations of quantum information science. While further research must be directed to the understanding of the emitter's quantum-optical properties under this fast excitation, hopefully leading to further improvements of the photon-indistinguishability, our results underline the considerable potential of fiber-pigtailed Purcell-enhanced sub-Poissonian quantum light sources for reaching clock-rates in implementations of quantum information comparable to laser-based systems.

\textit{Note added in proof}. - During peer-review of our manuscript, related work by Margaria \textit{et al.} appeared on arXiv reporting about a fiber-pigtailed micropillar-based single photon source~\cite{margaria_efficient_2024}.

\section{Acknowledgments}
We thank Maja Wasiluk for assistance with the optical characterization of the fiber-coupled device. 

\section{Data and Availability}
The data presented in this work is available from the authors upon reasonable request.

\section{Funding}
The authors acknowledge financial support by the German Federal Ministry of Education and Research (BMBF) via the project “QuSecure” (Grant No. 13N14876) within the funding program Photonic Research Germany, the BMBF joint project “tubLAN Q.0” (Grant No. 16KISQ087K), and by the Einstein Foundation Berlin via the Einstein Research Unit “Quantum Devices”. H.L., S.L., H.N., and Z.N. further acknowledge funding by the National Key Technology R\&D program of China (Grant No. 2018YFA0306101). S.Rodt and S.Reitzenstein acknowledge funding by the BMBF via the project QR.X Quantenrepeater.Link (Grant No. 16KISQ014) and the German Research Foundation via project INST 131/795-1 FUGG.

\section{Disclosure}
The authors declare no conflict of interest.

\setcounter{figure}{0}
\renewcommand{\figurename}{Fig.}
\renewcommand{\thefigure}{S\arabic{figure}}

\setcounter{table}{0}
\renewcommand{\tablename}{Tab.}
\renewcommand{\thetable}{S\arabic{table}}

\onecolumn
\section*{Supplementary Information}
\appendix

\section*{S1: FEM Simulations of FC-QD-hCBGs}
\label{sec:si_FEM}

\begin{figure*}[ht]
    \center
	\includegraphics[width=0.65\textwidth]{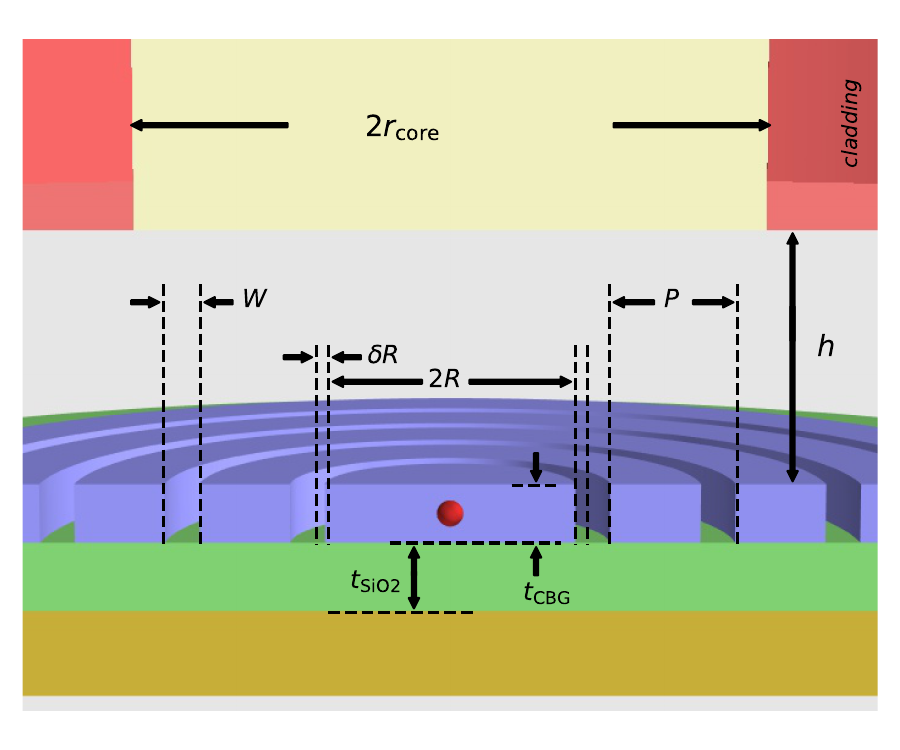}
	\caption{(a) Schematic depiction of a QD-hCBG cavity with a UHNA3 fiber exhibiting core-radius $r_\mathrm{core}$ laterally aligned to the hCBG's center and separated by distance $h$. The hCBG cavity of thickness $t_CBG$ consists of a central disc with radius $R+\delta{R}$ and a QD embedded in its center, surrounded by concentric rings of period $P$ and grating gap width $W$. The hCBG cavity is situated on a SiO$_2$ layer of thickness $t_{\mathrm{SiO}_2}$ on gold.}
\label{fig:figureS1}
\end{figure*}

\begin{table}[ht]
\begin{center}
\begin{tabular}{||c c c c c c c c||}
\hline
$R$ [nm] & $\delta{R}$ & $P$ [nm] & $W$ [nm] & $t_\mathrm{CBG}$ [nm] & $t_{\mathrm{SiO}_2}$ [nm] & m$_\mathrm{Rings}$ & $r_\mathrm{core}$ [nm] \\
\hline \hline
360 & 20 & 360 & 100 & 170 & 200 & 5 & 900 \\
\hline
\end{tabular}
\end{center}
\caption{hCBG cavity design parameters according to Fig.~\ref{fig:figureS1} and UHNA3 SMF dimensions. The parameter $m_\mathrm{Rings}$ represents the hCBG's number of rings.}
\label{tab:hCBG_parameters}
\end{table}

\begin{table}[ht]
\begin{center}
\begin{tabular}{||c c c c c||}
\hline
$n_\mathrm{GaAs}$ & $n_{\mathrm{SiO}_2}$ & $n_\mathrm{Au}$ & $n_\mathrm{core}$ & $n_\mathrm{clad}$\\
\hline \hline
3.460 & 1.450 & 0.12 + 6.33$i$ & 1.493 & 1.451\\
\hline
\end{tabular}
\end{center}
\caption{Refractive index parameters assumed in the FEM simulations for hCBG and UHNA3 SMF.}
\label{tab:indices}
\end{table}

\noindent The FEM simulations presented in this work were performed using the commercial FEM software JCMSuite (JCMwave GmbH, 2024). The QD emitter is simulated by a TE dipole source situated in the centre of the central hCBG disc in both vertical and lateral direction. The in-fiber efficiency is calculated as the power obtained from the overlap of
emitted CBG emission to the UHNA3-mode profile normalized to the total emitted dipole power. Further details on the simulation setup can be found in~\cite{rickert_optimized_2019}. The detailed hCBG-cavity design parameters and used fiber-core radius are listed in Table~\ref{tab:hCBG_parameters}, and the used refractive indices of the involved materials can be found in Table~\ref{tab:indices}.

\clearpage

\section*{S2: Influence of fiber distance on simulated Purcell enhancement}

\begin{figure*}[ht]
    \center
	\includegraphics[width=1\textwidth]{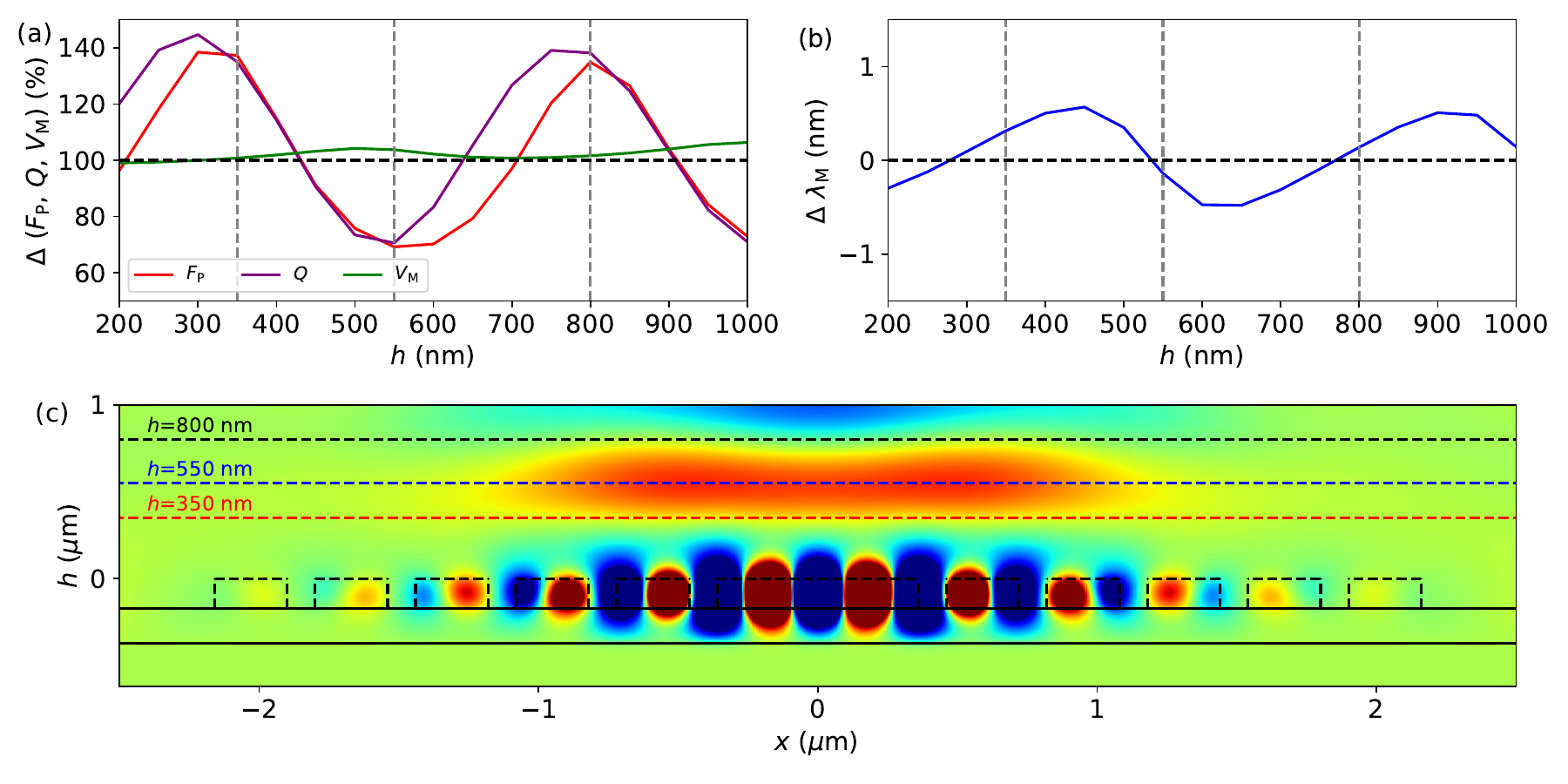}
	\caption{(a) Simulated relative Purcell factor $F_\mathrm{P}$, quality factor $Q$ and mode volume $V_\mathrm{M}$ for varying fiber-to-hCBG distance $h$. The values are given in percent relative to the respective values without the present fiber, indicated as dashed black line. $Q$ and $V_\mathrm{M}$ are obtained from eigenmode simulations, while $F_\mathrm{P}$ is obtained from scattering simulations at constant wavelength. (b) Simulated relative mode wavelength $\lambda_\mathrm{P}$ obtained from eigenmode simulations compared to the case without present fiber. (c) Field component $E_\mathrm{X}$ of the target hCBG mode without fiber present, obtained from eigenmode simulations. }
\label{fig:figureS2a}
\end{figure*}

As shown in Figure~1 in the main manuscript, the Purcell enhancement of a fiber-pigtailed QD-hCBG cavity depends noticeably on the fiber-to-cavity distance $h$. Figure~\ref{fig:figureS2a} provides further insights on how the presence of the fiber in close proximity to the hCBG cavity influences the cavity parameters.  
Figure~\ref{fig:figureS2a}(a) shows the $h$-dependent simulated Purcell factor $F_\mathrm{P}$, quality factor $Q$ and mode volume $V_\mathrm{M}$ relative to the respective values of the hCBG cavity without the present fiber. It is apparent, that $Q$ varies by up to $50$\% from the case without present fiber, while $V_\mathrm{M}$ stays nearly constant. It can therefore be assumed that the observed $h$ dependency of $F_\mathrm{P}$ originates almost exclusively from $Q(h)$. Note that the displayed simulation results were obtained from eigenmode simulations of the target hCBG mode in the case of $Q$ and $V_\mathrm{M}$, while $F_\mathrm{P}$ results from scattering simulations at fixed wavelength which takes also contributions of other modes into account. This is the reason for the slight offset of $Q$ and $F_\mathrm{P}$. It is further found that the mode wavelength exhibits small shifts dependent on $h$, as displayed in Fig.~\ref{fig:figureS2a}.

The simulated $Q(h)$-dependency can be understood if the emitted field intensity of the hCBG cavity is considered. Fig.~\ref{fig:figureS2a}(c) shows the $E_\mathrm{X}$ field component of the hCBG mode, which shows a near-field pattern with nodes and anti-nodes in vertical direction. If the fiber is placed at the position of minimum field intensity, as is for $h=350$\,nm, $Q$ exhibits a maximum, while $Q$ gets minimal if the fiber facet lies at positions of high field intensity, such as $h=550$\,nm. The reason is that an abrupt change in refractive index, such as the transition of air to fiber-core, is detrimental to mode confinement, a concept that was for example used as "gentle-confinement-designs" for high $Q$ photonic crystal cavities~\cite{akahane_high-q_2003,tanaka_design_2008}.

We note that the vertical modulation of the hCBG mode's electric field is present without the fiber present. If a fiber is brought in close proximity in this way, the hCBG's mode confinement is altered by the induced effective refractive index changes for the given field distribution. The fiber adds additional vertical confinement, effectively becoming part of the hCBG cavity. The predicted noticeable influence of the hCBG mode's $Q$ factor by the simulations renders this a versatile method to determine the fiber-to-cavity distance for fabricated structures in the experiment.

Note that Fig.~\ref{fig:figureS2a}(b) also predicts an $h$-dependent wavelength shift of the cavity mode of about $\pm0.5$\,nm. The shifts between the cooldowns of the pigtailed QD-hCBG device shown in Fig.~2(c) are however larger than expected from the simulations, indicating that the built-up strain might cause a change of the refractive index as mentioned in the main manuscript.

\clearpage

\section*{S3: Experimental setup}

\begin{figure*}[ht]
    \center
	\includegraphics[width=0.8\textwidth]{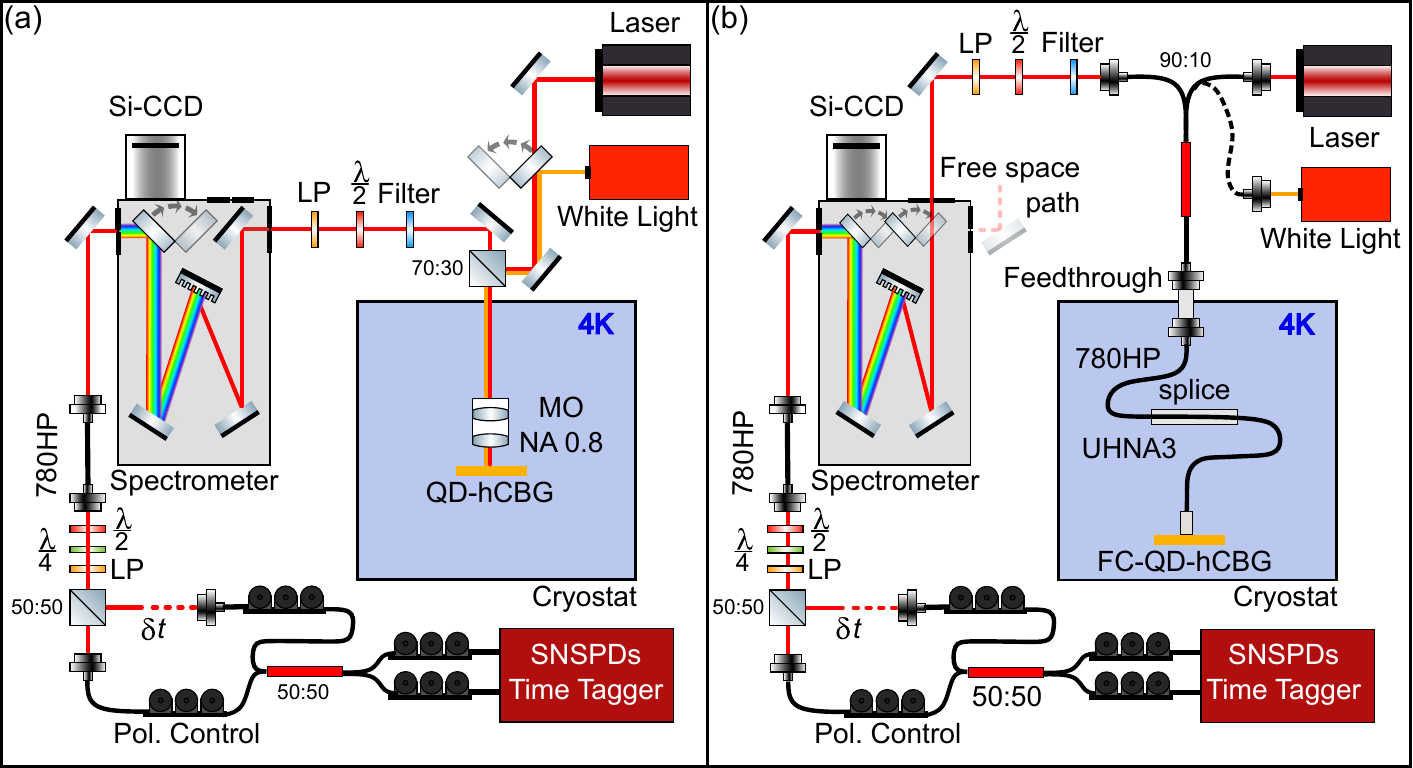}
	\caption{(a) Schematic depiction of the experimental setup used for free-space characterization of the hCBG cavity before the fiber pigtailing and (b) for the FC-device after the fiber-pigtailing. MO: microscope objective. LP: linear polarizer}
\label{fig:figureS2}
\end{figure*}

Figure~\ref{fig:figureS2} shows schematics of the experimental setup used for the quantum-optical measurements of the sample before and after the fiber-pigtailing. For the measurements before the pigtailing in the setup configuration in (a), the QD-hCBG cavity sample is placed in a closed-cycle helium cryostat and cooled to 4\,K with a low-temperature compatible microscope objective with NA=0.8 for confocal excitation/emission experiments. Excitation light reaching the sample via the excitation path is either from a pulsed 80\,MHz laser system with nominal 2\,ps temporal pulse-width (picoEmerald, APE GmbH) or from a broadband white-light source (SLS201L/M, Thorlabs GmbH). The emission from the sample exits the cryostat towards a 0.75\,m spectrograph (entrance slit width: 100\,$\mu$m) with 1200\,g/mm grating, using a half-wave-plate before the monochromator to match the polarization to the grating. For p-shell experiments, the emission from the sample is filtered with a bandpass filter ($\lambda_\mathrm{center}=(940 \pm 5)$\,nm) before entering the monochromator. To spectrally resolve the signal, the emission is reflected from the grating onto a 1340~pixel Si-CCD camera cooled to 200\,K with a Peltier element. For time-resolved measurements, the emission is reflected from the grating trough a separate spectrograph's exit (exit slit width: 80~$\mu$m) into a single mode fiber (SMF 780HP). We expect the spectral filtering caused by the grating and SMF in-coupling to be around 100\,$\mu$eV (0.07\,nm).
The SMF leads to a Hong-Ou-Mandel setup set as follows: An initial free-space polarization control consisting of half-waveplate, quarter-waveplate and linear polarizer ensures equal splitting at a polarizing BS to match both arms of the Mach-Zehnder interferometer (MZI) in intensity. The arms of the MZI are precisely time-matched for the time-difference $\delta{t}$ ($\delta{t}=2.0$\,ns and 12.5\,ns for 80\,MHz excitation rate, and $\delta{t}=781$\,ps for 1.28\,GHz excitation rate). The interference then occurs in a 50:50 fiber-beamsplitter, where the input polarizations are fully-controlled by polarization paddles in fiber to set the co- and cross-polarized interference. The output of the 50:50 fiber-beamsplitter leads to superconducting nanowire single photon detectors (SNSPDs) (Single Quantum EOS CS, Single Quantum B.V.) connected to time tagging electronics (QuTag, qtools GmbH). Additional polarization paddles on the fibers leading to the detector match the polarization to the nanowire orientation. The correct setting of polarization and time delay in the MZI is confirmed by interfering laser pulses at the QD emission wavelength and observing close to 100$\%$ interference contrast.

For the measurements of the FC-device, the sample was placed in the cryostat and the 780HP fiber output of the UHNA3-780HP fiber patchcord was connected to the cryostats vacuum fiber-feedtrough. On the room temperature side of the feedtrough, a 90:10 fiber-based beamsplitter was connected, whose 10\% input arm was used to connect the above mentioned 80\,MHz pulsed laser for optical excitation, or the broadband white-light source for reflection measurements. The 90\% fiber-beamsplitter output was connected to a free-space out-coupler towards the spectrometer. The same spectrometer and Si-CCD was used as for the free-space measurements, but was entered from an alternative entrance.

The efficiency estimation for the FC setup is discussed in S.I., section S8.

\clearpage

\section*{S4: Fiber-pigtailed spectra during cooldown}

\begin{figure*}[ht]
    \center
	\includegraphics[width=0.85\textwidth]{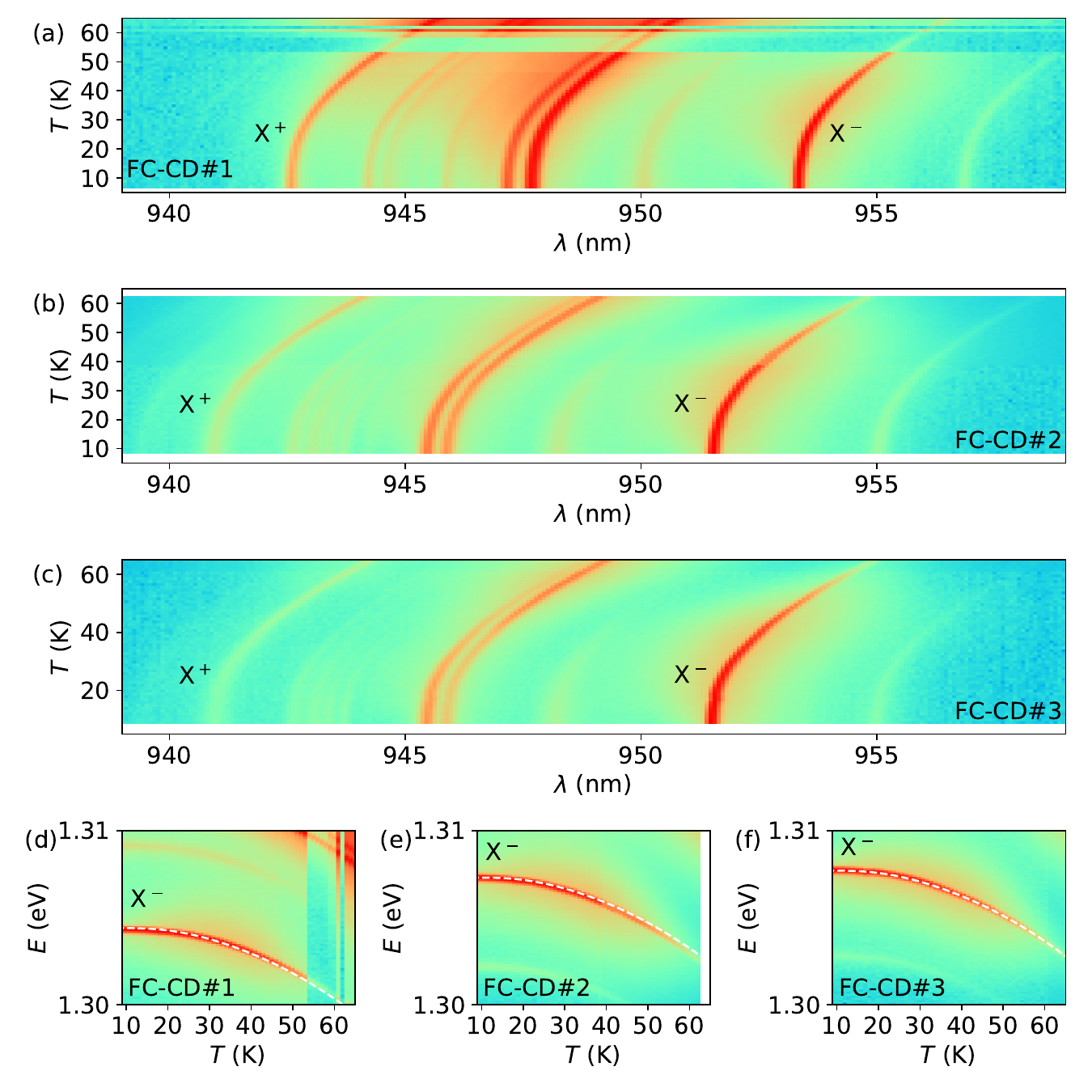}
	\caption{Recorded PL-spectra under above-band excitation during the cooldowns (a) \#1, (b) \#2 and (c) \#3 in logarithmic scale. The intensity variations in (a) at temperatures $>$54\,K stem from changes in the excitation power to check for saturation. (d)-(f) show the X$^-$ emission, respectively, with a dashed line following equation~(\ref{eq:Eg_vs_T}).}
\label{fig:figureS3}
\end{figure*}

\noindent Figure~\ref{fig:figureS3} shows the temperature dependent spectra during the three cooldowns of the FC-device for an excitation wavelength of $\lambda_\mathrm{exc}=793$\,nm. For cooldown \#1, the excitation power conditions at the QD were unknown, and the excitation power and integration time on the CCD was adjusted in the range above 54\,K to check for QD state saturation. The changes in intensity above 54\,K stem from these excitation power variations, and not from the sample itself. Once the cooldown was started, it could not be stopped until base temperature was reached, causing the intensity fluctuations to be recorded.

Fig.~\ref{fig:figureS3}(d)-(f) show the temperature-dependent emission of the X$^-$ line as a close-up for the respective cooldowns. The dotted line follows equation~(\ref{eq:Eg_vs_T}) for a temperature dependent band gap energy $E_\mathrm{g}$ of GaAs (shifted by a set confinement energy for each cooldown), with the corresponding parameters listed in~\cite{passler_temperature_2001}:
\begin{equation}
    \label{eq:Eg_vs_T}
    E_\mathrm{g} = E_0 -\alpha\left(\dfrac{W_1\Theta_1}{\exp(\Theta_1/T)-1} + \dfrac{W_2\Theta_2}{\exp(\Theta_2/T)-1} \right).
\end{equation}

The good agreement with theory indicates that the FC-device reaches the indicated temperatures (ultimately 4.8\,K), and the wavelength shifts compared to the sample before pigtailing cannot be explained by temperature, as elaborated in the main text. We note further that equation~(\ref{eq:Eg_vs_T}) with InAs parameters does not fit the experimental data well, indicating that the QD's temperature dependence is either largely determined by the surrounding GaAs matrix material, or significant Ga incorporation takes place during the growth.

\clearpage

\section*{S5: Assigment of QD states to emission lines.}

\begin{figure*}[ht]
    \center
	\includegraphics[width=1\textwidth]{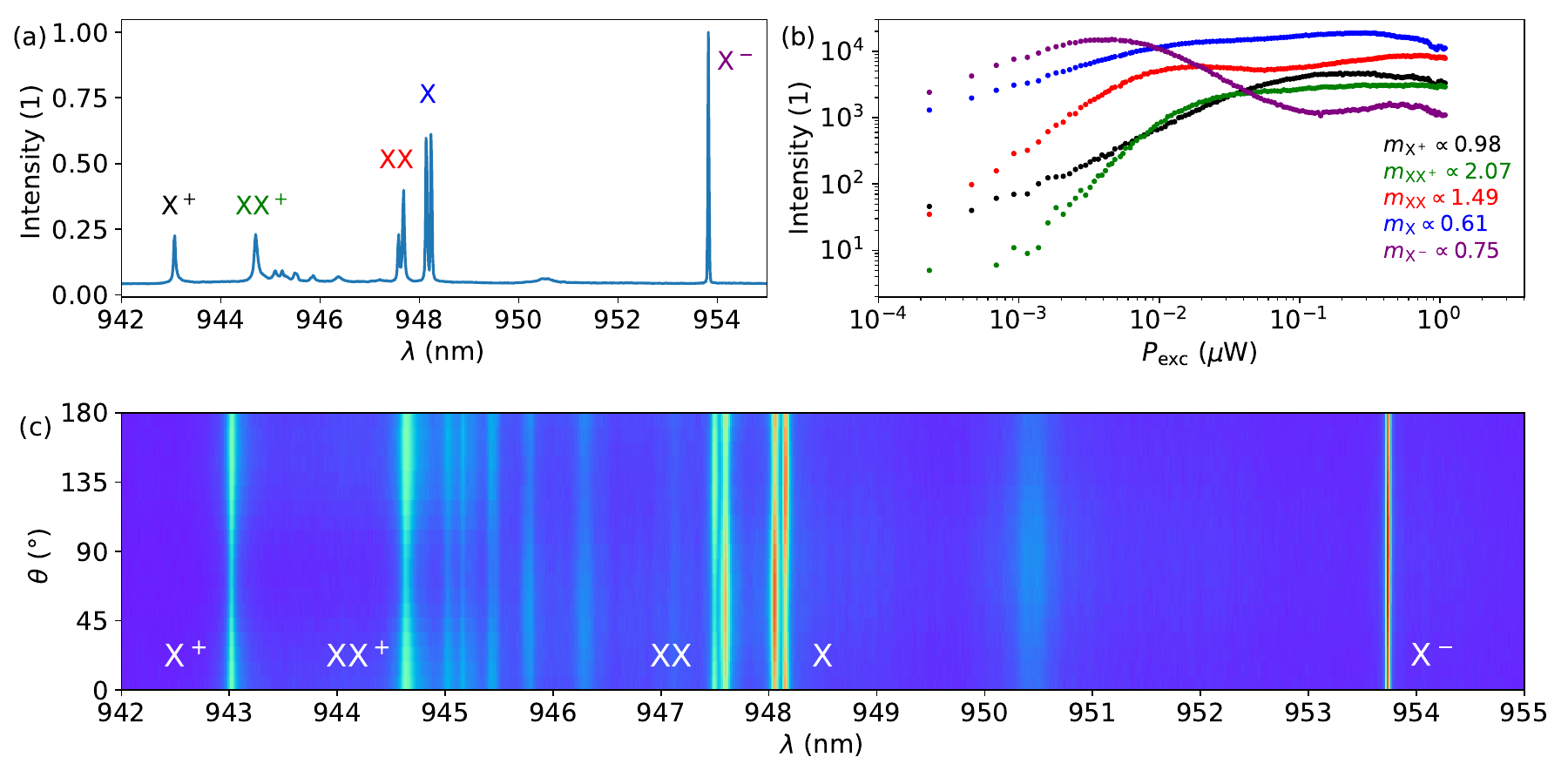}
	\caption{\textbf{Excitation power series and polarization series measurements to identify the QD states of the fiber-pigtailed QD-hCBG device}. The measurements were obtaiend for Cooldown \#1 under off-resonant excitation with $\lambda_\mathrm{exc}=793$\,nm at $T=4.8$\,K. (a) Spectrum of the fiber-pigtailed QD-hCBG device with assigned QD-states. (b) Peak intensity of the labeled states in (a) for varying excitation powers $P_\mathrm{exc}$. The slope $m$ obtained from linear fits of the data in this double-logarithmic scale is listed. (c) Polarization resolved spectra under off-resonant excitation for detection angles $\theta$ with indicated QD states.}
	\label{fig:S4_Pol_and_PowerSeries}
\end{figure*}

\noindent To assign the observed emission lines in the spectrum of the fiber-pigtailed QD-hCBG device excitation power dependent and detection polarization dependent measurements were conducted during during the first cooldown. The data is displayed in Figure~\ref{fig:S4_Pol_and_PowerSeries}, with the assigned states for the spectrum shown in Fig.~\ref{fig:S4_Pol_and_PowerSeries}(a).

The XX and X emission lines are assigned by the visible fine-structure splitting in the polarization resolved detected emission intensity in Fig.~\ref{fig:S4_Pol_and_PowerSeries}(c), and X is distinguished from XX based on the lower slope from the excitation power dependent measurements in Fig.~\ref{fig:S4_Pol_and_PowerSeries}(b). 

Neither of the assigned X$^+$, XX$^+$ and X$^-$ states show a (resolvable) fine-structure splitting for varying polarization, indicating trionic states. X$^+$ and XX$^+$ are distinguished based on their power dependency. The appearance of X$^+$ and XX$^+$ at shorter wavelength (i.e. anti-binding) compared to the X is commonly observed for InAs QDs, while the negative trion appears at longer wavelengths (i.e. binding) compared to X~\protect\cite{rodt_correlation_2005}.

\clearpage

\section*{S6: Off-resonant spectra before and after fiber-pigtailing}

\begin{figure*}[ht]
    \center
	\includegraphics[width=0.85\textwidth]{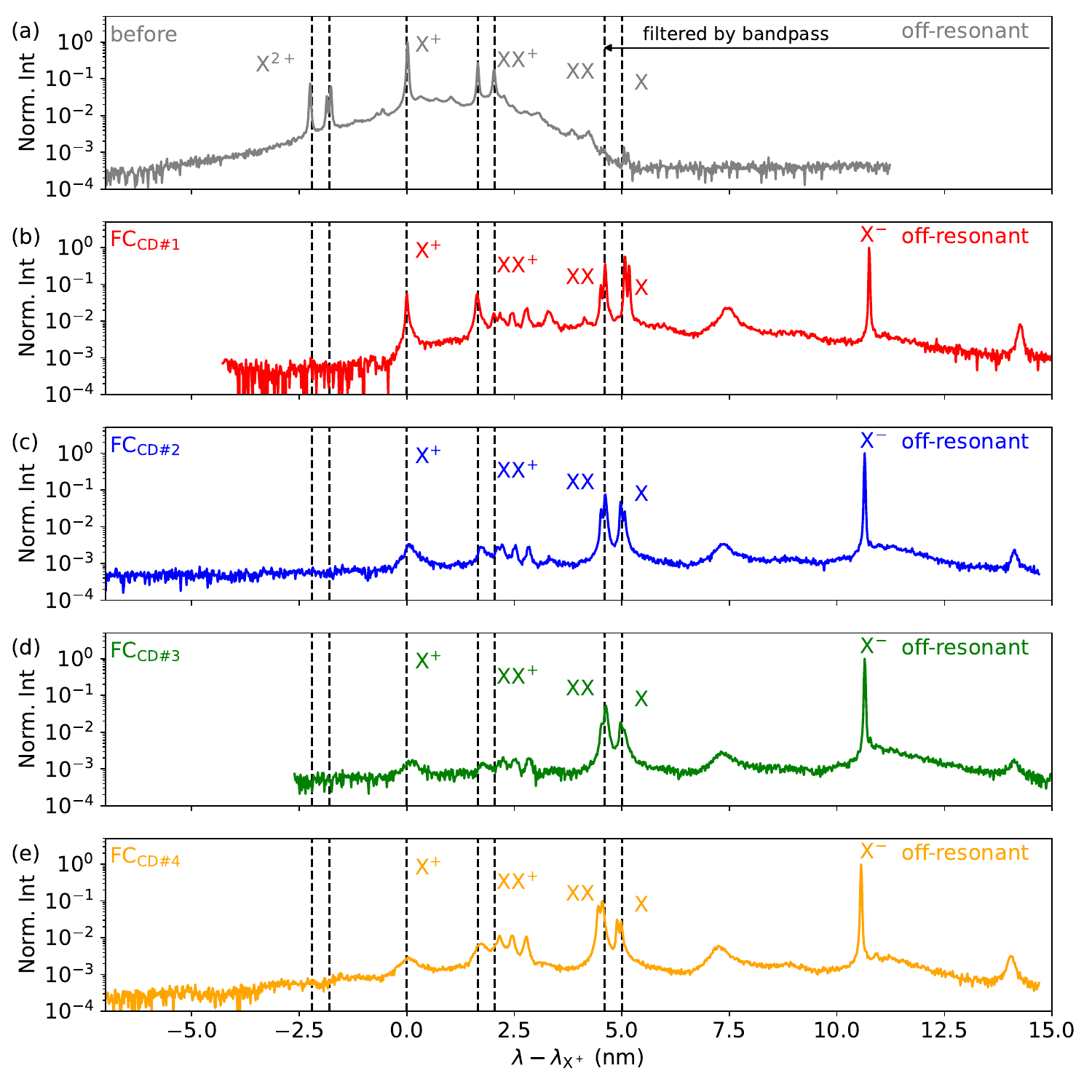}
	\caption{PL-spectra under above-band ($\lambda_\mathrm{exc}=793$\,nm) excitation (a) before and (b-e) after the fiber-pigtailing for the respective cooldowns. The normalized intensity is displayed in logarithmic scale. The recorded emission wavelength is normalized to the emission wavelength of the X$^+$. The spectral positions of associated QD states are indicated as dashed lines. For the measurement before the pigtailing, the filtered-out spectral region caused by the bandpass filter is indicated.}
\label{fig:figureS4}
\end{figure*}

\noindent
Figure~\ref{fig:figureS4} shows the spectra in Fig. 2(a) in the main text under off-resonant excitation ($\lambda_\mathrm{exc}=793$\,nm) before and after the fiber-pigtailing plotted in logarithmic scale. The wavelength on the x-axis is normalized to the emission wavlength of the X$^+$ transition. Respective QD states corresponding to emission lines are indicated, as well as the filtered-out spectral region by the employed bandpass filter for the spectrum before the coupling.

X$^+$, XX$+$, XX and X lines appear at identical spectral positions, clearly proving that the fiber-pigtailing was deterministic for the pre-selected QD-hCBG cavity. The strong emission background in the spectrum before the pigtailing in Fig.~\ref{fig:figureS4}(a) is caused by $\sim$ 10$\times$ higher excitation power $P_\mathrm{exc}$. Similarly, $P_\mathrm{exc}$ was slightly higher for cooldowns \#1 and \#4 for the displayed spectra, causing slightly varying backgrounds.
An excition wavlength-dependent comparison before and after the pigtailing is discussed further in S.I., section~S7.

\clearpage

\section*{S7: Photoluminescence excitation  before and after fiber-pigtailing}

\begin{figure*}[ht]
    \center
	\includegraphics[width=0.55\textwidth]{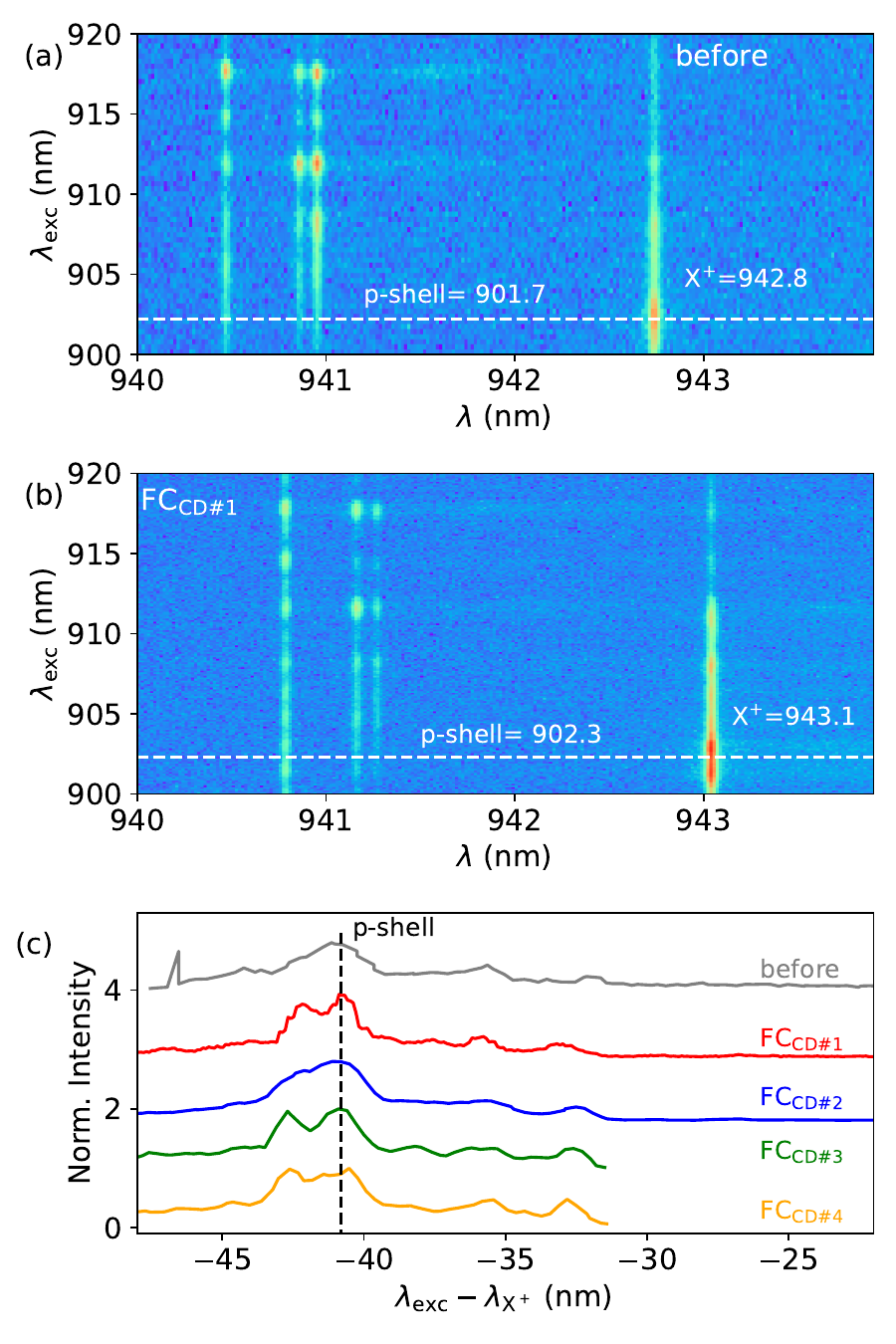}
	\caption{Photoluminescence excitation (PLE) scans for varying excitation wavelengths $\lambda_\mathrm{exc}$ and corresponding emission wavelength $\lambda$. (a) PLE before and (b) after the fiber-pigtailing for cooldown \#1. The respective p-shell is indicated. (c) Intensity of the X$^+$ emission line for PLE scans before and after the pigtailing for respective cooldowns with indicated p-shells.}
\label{fig:figureS4b}
\end{figure*}

\noindent
Figure~\ref{fig:figureS4b} shows photoluminescence excitation (PLE) data before and after the pigtailing, with detailed PLE scans before in Fig.~\ref{fig:figureS4b}(a) and after the pigtailing for cooldown $\#1$ in in Fig.~\ref{fig:figureS4b}(b). The respective p-shell and X$^+$ wavelengths are indicated. Fig.~\ref{fig:figureS4b}(c) shows the intensity of the X$^+$ line for varying excitation wavelengths normalized to the trion emission wavelengths, i.e. the s-p splitting. The fact that identical s-p splitting is observed before and after the pigtailing proves the deterministic fiber-pigtailing technique.

\clearpage

\section*{S8: Setup efficiency estimation}

\begin{figure*}[ht]
    \center
	\includegraphics[width=0.625\textwidth]{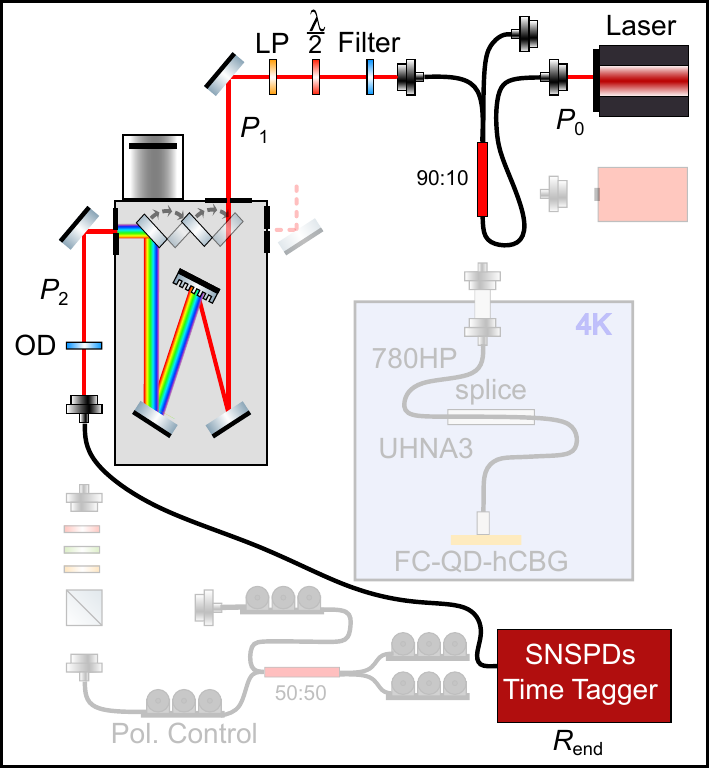}
	\caption{Schematic depiction of the measurement to estimate the efficiency of the setup for the FC-device.}
\label{fig:figureS5}
\end{figure*}

\noindent Figure~\ref{fig:figureS5} shows the measurement to estimate the setup efficiency for the characterisation of the fiber-pigtailed sample. The fiber-based beamsplitter's input, which would be connected to the cryostat's fiber-feedthrough for the characterization of the FC-device, is connected to a continuous-wave (cw) laser at $\lambda_{\mathrm{X}^+}$ with input power $P_0 = 5\,\mu$W. The 10\% output of the fiber-based beamsplitter is blocked, and the cw-laser signal reaches the spectrometer via the 90\% output. The power $P_1$ is measured at the entrance of the spectrometer. The power $P_2$ is the cw-laser power at the exit of the spectrometer, and is attenuated with OD-filters before entering the fiber towards the SNSPD detectors. $R_\mathrm{end}$ is the observed countrate on the SNSPDs.

We find $P_1/P_0$ = 0.327, $P_2/P_1$ = 0.333 and $(R_\mathrm{end}hc/\lambda_{\mathrm{X}^+})/P_2$ = 0.45. This leads to an overall setup efficiency from the fiber-based beamsplitter input to the detectors of 0.049. For the experimental in-fiber efficiency at the UHNA3 fiber-facet, the efficiency by the splice (0.95) and by the vacuum fiber-feedthrough (0.6) have to be additionally taken into account, as described in the main text.

\clearpage

\section*{S9: Influence of $T_1$ on $V_\mathrm{HOM}$}
\label{sec:si_VHOMvsT1}

\begin{figure*}[ht]
    \center
	\includegraphics[width=0.7\textwidth]{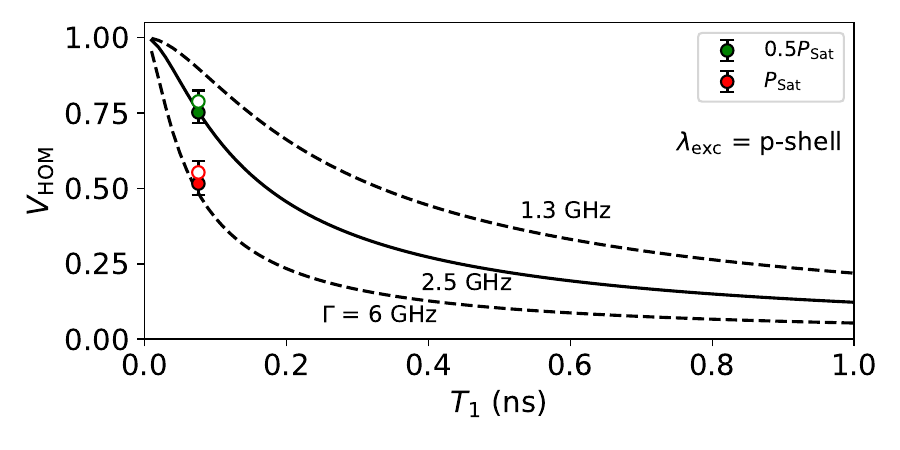}
	\caption{Measured HOM visibilities $V_\mathrm{HOM}$ at $\delta{t}=12.5$\,ns for observed $T_1$-time of the fiber-pigtailed device and varying fraction of the saturation power $P_\mathrm{sat}$. The indicated lines correspond to equation~(\ref{eq:VHOM_vs_T1}) for  different Gaussian line-broadening $\Gamma$.}
\label{fig:figureS6}
\end{figure*}

\noindent 
The following S.I. provides a reference for the measured two-photon-indistinguishablity $V_\mathrm{HOM}$ values in p-shell excitation for the fiber-pigtailed device, since we did not measure $V_\mathrm{HOM}$ before the pigtailing.
The $T_1$-influence on the indistinguishability of two emitted photons can be described as~\protect\cite{gold_two-photon_2014, nawrath_resonance_2021}:
\begin{equation}
\label{eq:VHOM_vs_T1}
	V_\mathrm{HOM}(T_1,\Gamma) = \dfrac{A(\Gamma)}{T_1}\exp\left({\dfrac{A(\Gamma)^2}{\pi{T_1^2}}}\right)\cdot\mathrm{erfc}\left(\dfrac{A(\Gamma)}{\sqrt{\pi}T_1}\right)
\end{equation}
with the complementary error-function $\mathrm{erfc()}$, factor $A(\Gamma) = \frac{\sqrt{\ln{2}}}{\sqrt{2\pi}\Gamma}$, and the inhomogeneously broadened QD emission line following a Gaussian distribution with FWHM-linewidth $\Gamma$. Figure~\ref{fig:figureS6} shows the measured $V_\mathrm{HOM}$ values of the pigtailed QD-hCBG device for p-shell excitation at different fractions of $P_\mathrm{sat}$ discussed in the main text plotted against the observed $T_1$-time.
The results of equation~(\ref{eq:VHOM_vs_T1}) are shown for $\Gamma$-values between 1.3\,GHz and 6\,GHz (i.e., linewidths of 0.005\,nm and 0.028\,nm, respectively), in agreement with inhomogenous-broadening values previously reported for InAs QD-hCBG cavities under p-shell excitation~\protect\cite{rickert_high_2024}.
The photon-indistinguishability observed for the fiber-pigtailed device, as discussed in the main text, are hence very similar to free-space operated QD-hCBG cavities. It can be expected, that even higher indistinguishabilities under non-resonant excitation can be reached, if $T_1$ can be further reduced.

\clearpage

\section*{S10: Additional data from a second fiber-pigtailed QD-hCBG device}
\label{sec:si_FCF13}

\begin{figure*}[h]
    \center
	\includegraphics[width=0.93\textwidth]{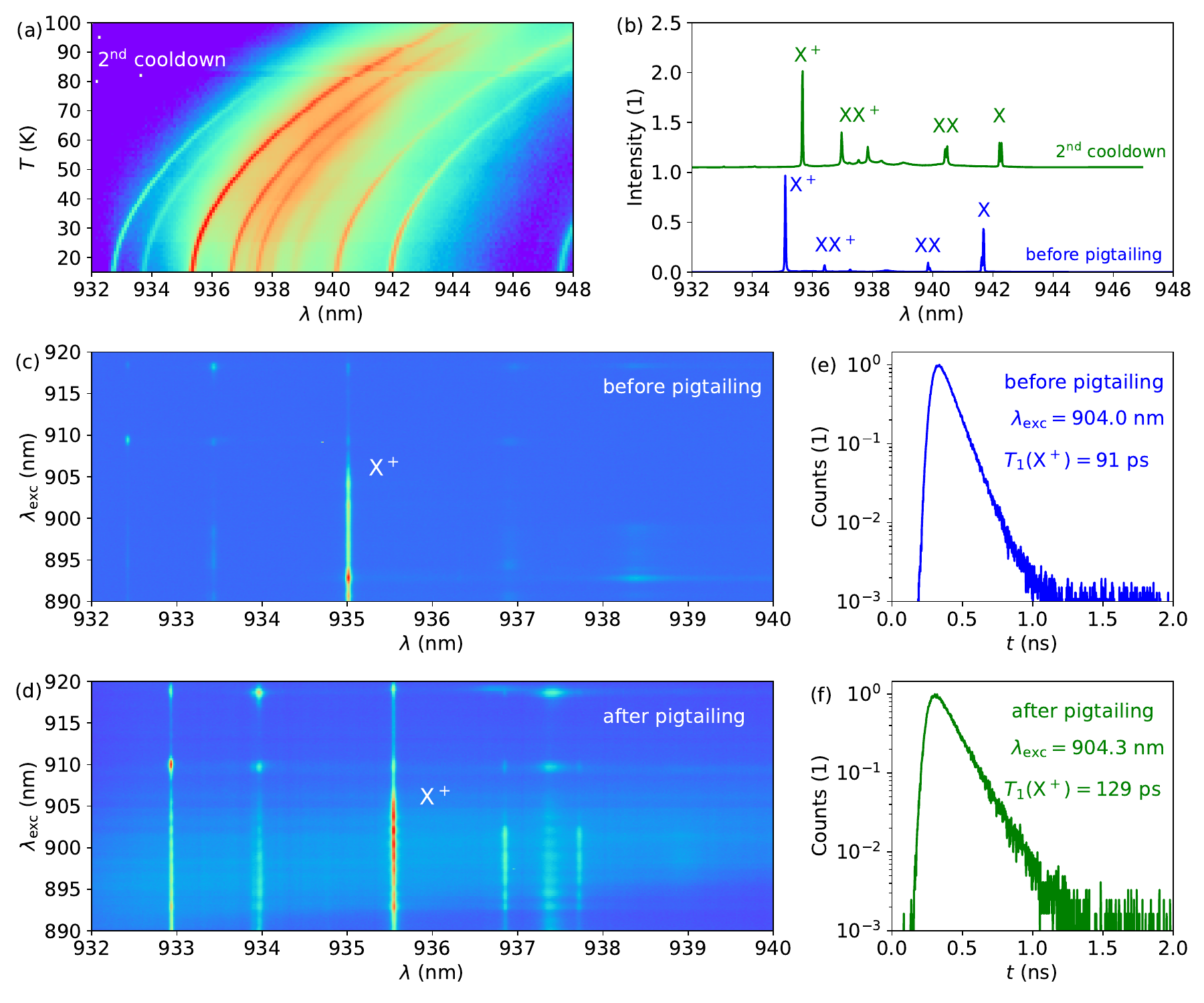}
	\caption{\textbf{Additional data for the second fiber pigtailed QD-hCBG device FC-CBG \#2.} (a) Emission spectra under $\lambda_\mathrm{exc}=793$\,nm excitation during the second cooldown after the pigtailing. (b) Emission spectra at base temperature before and after the pigtailing. (c) PLE spectra before and (d) after the pigtailing. (e) $T_1$-time of X$^+$ for p-shell excitation before and (f) after the fiber-pigtailing.}
	\label{fig:FCF13}
\end{figure*}

We performed a fiber-pigtailing process for a second sample. In the following, we will refer to this second pigtailed as FC-CBG \#2.
In contrast to the discussed pigtailed device in the main manuscript, device FC-CBG \#2 showed a sudden decrease in emission intensity a couple of minutes after reaching base temperature in a first cooldown. The emission properties observed during a second cooldown are displayed in Figure~\ref{fig:FCF13}.   

Figure~\ref{fig:FCF13}(a) shows recorded spectra under off-resonant excitation during the second cooldown. Figure~\ref{fig:FCF13}(b) shows the device's emission at $T=4.8$\,K in green, as well as the spectrum before the pigtailing in blue.
Despite the reduced intensity after the first cooldown, the fiber-to-cavity alignment was not completely lost, and further spectra, photoluminescence excitation (PLE) scans and time-resolved measurements could be conducted during the second cooldown, as displayed in Figure~\ref{fig:FCF13}(c-f). Based on the observed quasi-resonances in the PLE and the spectral positions of the assigned QD states, the pigtailed QD-hCBG device is unambiguously identified as the target device before the pigtailing, albeit again with a slight red shift.
The measured $T_1$-time of X$^+$ appears to be once again slower than before the pigtailing. Note that for this device FC-CBG \#2, it was not possible to obtain information on the cavity mode from reflection measurements, since the signal-to-noise ratio was too low stemming from the fiber-displacement, so no information can be gained on the mode's $Q$-factor after the pigtailing.

To put this observed properties of device FC-CBG \#2 in perspective with the pigtailed device in the main manuscript: we believe that the pigtailed device discussed in the manuscript experienced an elastic deformation and strain built-up during the first thermal cycle, which appears to equilibrated after the first cooldown. 
For device FC-CBG \#2, the thermal stressed appeared to be enough to cause a plastic deformation of the fiber-to-hCBG connection, which resulted in the observed deteriorated alignment and loss in emission intensity. The plastic deformation however, is also believed to have caused a partial relaxation of the strain accumulated during the first cooldown. This can explain why there are no intensity shifts observed in this second sample, compared to the device discussed in the manuscript.

The reason why device FC-CBG \#2 showed a plastic deformation, while the device discussed in the manuscript had intact fiber-alignment after the first cooldown could originate from several factors: Firstly, the temperature ramp of the cryostat during 1$^\mathrm{st}$ cooldown of device FC-CBG \#2 was steeper than for the second cryostat used for the later cooldowns, potentially causing higher thermal stress. Secondly, we think that the membrane area being in contact with the fiber might play a crucial part in the strain distribution and potential deformations. 

\twocolumn

\bibliographystyle{ieeetr}
\bibliography{bibliography_FC}


\end{document}